\documentclass[journal]{IEEEtran}
\makeatletter

\newcommand{\Rmnum}[1]{\expandafter\@slowromancap\romannumeral #1@}
\makeatother

\usepackage{mathrsfs}
\renewcommand{\IEEEQED}{\IEEEQEDopen}

\usepackage{ulem} 

\usepackage{stfloats}
\usepackage{dsfont}
\usepackage{cite}
\usepackage{graphicx}
\usepackage{subfigure}
\ifCLASSINFOpdf

\else

\fi
\usepackage[cmex10]{amsmath}
\usepackage{float}
\usepackage{array}
\usepackage{algorithm}  
\usepackage{algorithmic}  
\usepackage{amsmath}
\usepackage{amsfonts}
\usepackage{amssymb}
\usepackage{ulem}
\usepackage{cancel}
\newcommand{\rev}{\textcolor {black}}

\usepackage{xcolor}
\begin{document}
%

\title{Signal Detection in MIMO Systems with Hardware Imperfections: Message Passing on Neural Networks}

%
%
\author{Dawei Gao, 
       Qinghua Guo, 
       Guisheng Liao, 
       Yonina C. Eldar, ~\IEEEmembership{Fellow, IEEE},
       Yonghui Li, ~\IEEEmembership{Fellow, IEEE}, 
       Yanguang Yu, 
     and Branka Vucetic,  ~\IEEEmembership{Fellow, IEEE}

\thanks{Corresponding to Qinghua Guo (qguo@uow.edu.au).}
\thanks{Dawei Gao and Guisheng Liao are with the Hangzhou Institute of Technology, Xidian University, Hangzhou 311200, China and also with the National Laboratory of Radar Signal Processing, Xidian University, Xi'an 710071, China
(e-mail: gaodawei@xidian.edu.cn; liaogs@xidian.edu.cn).}
\thanks {Qinghua Guo and Yanguang Yu are with the School of Electrical, Computer and Telecommunications Engineering, University of Wollongong, NSW 2522, Australia (e-mail: qguo@uow.edu.au; yanguang@uow.edu.au).}
\thanks{Yonina C. Eldar is with the Faculty of Math and CS, Weizmann Institute of
Science, Rehovot, 7610001, Israel (email: yonina.eldar@weizmann.ac.il).}
\thanks{Yonghui Li and Brank Vucetic are with the School of Electrical and Information Engineering,
University of Sydney, Sydney, NSW 2006, Australia (e-mail: yonghui.li@
sydney.edu.au; branka.vucetic@sydney.edu.au).}

}

%
%

\markboth{}%
{Shell \MakeLowercase{\textit{et al.}}: Bare Demo of IEEEtran.cls for IEEE Communications Society Journals}
%



\maketitle


\begin{abstract} 
In this paper, we investigate signal detection in  
multiple-input-multiple-output (MIMO) communication systems with hardware impairments, such as power amplifier nonlinearity and in-phase/quadrature imbalance.
To deal with the complex combined effects of hardware imperfections, neural network (NN) techniques, in particular deep neural networks (DNNs), have been studied 
to directly compensate for the impact of hardware impairments. However, it is difficult to train a DNN with limited pilot signals, hindering its practical applications. \rev{In this work, we investigate how to achieve efficient Bayesian signal detection in MIMO systems with hardware imperfections. Characterizing combined hardware imperfections often leads to complicated signal models, making Bayesian signal detection challenging. To address this issue, we first train an NN to `model' the MIMO system with hardware imperfections and then perform Bayesian inference based on the trained NN.} 
\rev{Modelling the MIMO system with NN
enables the design of NN architectures based on the signal flow of the MIMO system, minimizing the number of NN layers and parameters, which is crucial to achieving efficient training with limited pilot signals. We then represent the trained NN with a factor graph, and design an efficient message passing based Bayesian signal detector, leveraging the unitary approximate message passing (UAMP) algorithm.} The implementation of a turbo receiver with the proposed Bayesian detector is also investigated. Extensive simulation results demonstrate that the proposed technique delivers remarkably better performance than state-of-the-art methods.       
\end{abstract}

\begin{IEEEkeywords}
Hardware imperfections, I/Q Imbalance, power amplifier nonlinearity, multiple-input-multiple-output (MIMO), neural networks (NNs), factor graphs, approximate message passing (AMP), Bayesian inference.  
\end{IEEEkeywords}

%
\IEEEpeerreviewmaketitle

\section{Introduction}
\IEEEPARstart{W}{e} consider signal detection for  multiple-input multi-output (MIMO) communications in the presence of hardware impairments, which arise, e.g., in millimeter wave (mm-wave) communications, where mm-wave front ends suffer from significant hardware imperfections, compromising signal transmission quality and degrading system performance {\cite{7400949,8094882,8333742}}.
{A pronounced impairment is in-phase/quadrature (I/Q)
imbalance, i.e., the mismatch of amplitude,
phase and frequency response between the I and Q branches, which impairs their orthogonality {\cite{9096604}}. Power amplifier (PA) nonlinearity leads to nonlinear distortions to transmitted signals, which cannot be overlooked, especially in mm-wave communications \cite{8094882}.
The hardware imperfections need to be handled properly to avoid inducing significant system performance loss.}

{
{Many techniques have been considered to mitigate the impact of hardware imperfections.} To handle PA nonlinearity, Volterra series based techniques were proposed for nonlinearity compensation at either transmitter or receiver} \cite{552219,6353238}. However, these techniques often need to determine a large number of Volterra series coefficients, which is a difficult task. To address this, some simplified methods such as those based on memory polynomials \cite{1264205}, Hammerstein model \cite{5745202} and Wiener model \cite{5259211} were proposed \cite{1264205}. {Addressing I/Q imbalance has also attracted much attention} \cite{9004558,4378242,4773160,9129370}. {In \cite{4773160}, a
dual-input nonlinear model based on a real-valued Volterra series
was proposed to model the I/Q imbalance, and its inverse model was employed {at the transmitter} to pre-compensate the I/Q imbalance. 
{In \cite{9129370}, a single-user point-to-point mm-wave hybrid beamforming system with I/Q imbalance at the transmitter and its pre-compensation were
considered. {The pre-compensation technique \cite{9129370} assumes the availability of instantaneous channel state information at
the transmitter, which can be difficult to achieve in practical scenarios.}}} 
{{With higher orders, polynomial-based techniques have potential to handle severer nonlinear distortions, which, however, are more prone to numerical instability in determining their coefficients \cite{1337325, 7492233,8388717}.}}    
{We also note that, most of the polynomial-based algorithms in the literature deal with a single type of hardware imperfections, i.e., either PA nonlinearity or I/Q imbalance.} However, hardware imperfections may occur at the same time, leading to combined effects. 

{Neural networks (NNs) have recently emerged as a promising technique to deal with the nonlinear effects in communication systems{\cite{GAO2019102594,9310357,9292976}}}. {In \cite{1273746}, a real-valued time-delay neural network (RVTDNN) was proposed to model PA behaviors. 
Various variants of RVTDNN were proposed \cite{8383719,8476222} to address the combined effects of hardware impairments. 
{In \cite{8383719}, high-order signal components are applied to the RVTDNN to pre-compensate both the PA nonlinearity and I/Q imbalance. In \cite{8476222}, a deep NN (DNN) based technique was proposed to mitigate combined PA nonlinearity and I/Q imbalance at the transmitter of a MIMO system. 
{In \cite{9322327}, a residual NN was proposed for digital predistortion, where shortcut connections are added between the input and output layer to improve the performance of PA nonlinearity mitigation. These predistortion based methods require feedback from the receiver, which can be inconvenient or difficult to implement, especially in the case of time-variant environments.
Post-compensation techniques at the receiver have also been investigated \cite{8693029,9183972}. A recurrent NN (RNN) was proposed in \cite{8693029} to compensate PA nonlinearity in a fiber-optic link. In \cite{9183972}, a deep-learning (DL) framework that integrates feedforward NN (FNN) and RNN was proposed to combat both the nonlinear distortion and linear interference. However, these works do not consider the impact of I/Q imbalance. }}} 
Moreover, a significant problem with the DNN based techniques is that a large number of pilot symbols are required to train the DNNs properly, leading to unacceptable overhead and hindering their application especially in time-varying environments.  

In this work, we investigate the issue of signal detection in an uplink multi-user mm-wave MIMO system, where transmitters (at users) suffer from combined distortions of PA nonlinearity and I/Q imbalance due to the use of low-cost mobile devices. \rev{To combat the combined effects of hardware imperfections and multi-user interference, the conventional approach is to design a DNN based detector with received signal as input and predicated symbols as output (shown in Fig. 1), which we call direct detection. However, it is difficult to train the DNN with limited pilot symbols. Due to the superior performance of Bayesian signal detection, in this work, we investigate how to achieve efficient Bayesian detection in the presence of combined hardware imperfections. Bayesian detection relies on a signal model. However, characterizing combined hardware
imperfections in a MIMO system leads to a complicated signal model
(which may also be subject to modelling errors), making Bayesian signal detection
challenging. We propose a new strategy, where we first use an NN to `model' the MIMO system (i.e., the NN serves as a substitute for the signal model), which captures combined effects of hardware imperfections and multi-user interference. Then we perform Bayesian inference based on the trained NN. We call this indirect detection.  
This strategy enables us to design the NN architecture based on the signal flow of the MIMO system and minimize the number of layers and parameters of the NN, making it possible to achieve efficient training with limited pilot symbols.}  

\rev{To perform Bayesian inference with the trained NN, we represent it with a factor graph and develop message passing based Bayesian signal detection. The presence of densely connected factors due to the NN weight matrices makes the Bayesian inference difficult.}  
The approximate message passing (AMP) algorithm is promising in handling densely connected factor graphs \cite{6875146}. 
However, AMP works well for i.i.d (sub-) Gaussian matrices, but suffers severe performance degradation or easily diverges for a general matrix\cite{6875146}. The work in \cite{2015arXiv150404799G} shows that AMP can still work well in the case of a general matrix when a unitary transform of the original model is used. The variant of AMP is called unitary AMP (UAMP), which was also known as UTAMP \cite{2015arXiv150404799G, 9547768, 9293406}. As NN weight matrices are normally not i.i.d. (sub-) Gaussian, we adopt UAMP and show that it plays a crucial role in achieving efficient message passing based Bayesian inference.        

The contributions of this work are summarized as follows:
\begin{itemize}
    \item {A new strategy to achieve Bayesian signal detection for a  communication system with complicated input-output relationship: We use an NN to model the behaviour of the MIMO system, followed by Bayesian inference based on the NN. This indirect detection strategy is more efficient than direct detection. Although this work focuses on MIMO systems with I/Q imbalance and PA nonlinearity, the developed method can be extended to deal with a general system with complicated input-output relationship.}
    \item Signal-flow-based NN architecture design: The architecture of the NN is carefully designed based on the signal flow of the MIMO system, so that the number of layers and parameters of the NN is minimized, which is crucial to achieving efficient training. 
    \item Message passing based Bayesian inference on NNs: To realize Bayesian signal detection based on an NN, we represent the NN as a factor graph and an efficient UAMP-based message passing inference algorithm (called MP-NN) is developed. 
    \item Iterative detection and decoding in coded systems: Another advantage of the new strategy is that the proposed MP-NN Bayesian detector is able to work with a soft-in-soft-out (SISO) decoder, leading to a much more powerful turbo receiver. In contrast, it is unknown how to develop a turbo receiver with existing DNN or polynomial based direct detection techniques.   
  \item Comparisons with existing techniques: We carry out various comparisons with state-of-the-art methods and demonstrate that the proposed approach delivers remarkably better performance. 
\end{itemize}

The remainder of the paper is organized as follows. In Section II, the signal model of MIMO communications with combined hardware imperfections is given and existing techniques are introduced. In Section III, with the new strategy, we investigate the NN architecture design and training, and develop a UAMP-based Bayesian detector by performing message passing on the trained NN.   
The extension to turbo receiver in a coded system is investigated in Section IV. Simulation results are provided in Section V, followed by conclusions in Section VI.
The notations used in this paper are as follows. Boldface lower-case and upper-case letters denote vectors and matrices, respectively. The superscript $(\cdot)^{*}$ represents the conjugate operation. The notations
$(\cdot)^T$ and $(\cdot)^H$ represent the transpose and conjugate transpose operations, respectively. We use $|x|$ and $||\mathbf{x}||$ to denote the amplitude of $x$ and the norm of $\mathbf{x}$, and use $\Re\{\cdot\}$ and $\Im\{\cdot\}$ to represent the real and imaginary parts of a complex number, respectively. {The notation$ \langle f(x)\rangle_{p(x)}$ denotes the expectation of $f(x)$ with respect to distribution $p(x)$}.

\section{Signal Model and Existing Methods}

\subsection{Signal Model}

We consider an uplink transmission of a multi-user mm-wave MIMO system with $K$ users. Considering the cost of mobile devices, we assume that each user has a single antenna, where low-cost modulators and PAs are used, resulting in I/Q imbalance and PA nonlinear distortions during transmission \cite{7600411}. The base station (BS) is equipped with $N$ antennas. 

{The $m$th symbol of user $k$ is denoted by $x_k(m)\in \mathcal{A}$, where $\mathcal{A}$ denotes the symbol alphabet. The symbols of all users at time instant $m$ form a vector $\mathbf{x}(m)$}. At the transmitter side, the signal is up-converted to radio frequency through modulation, and the mismatch between I and Q branches is characterized as \cite{8476222}  
\begin{equation} \label{newimbalance}
    {x}^{a}_k(m) = \xi_kx_k(m)+\zeta_kx^{*}_k(m),
\end{equation}
where 
\begin{eqnarray}
\xi_k = \cos(\frac{\theta_k}{2})+j\lambda_k \sin(\frac{\theta_k}{2}),\\
\zeta_k = \lambda_k \cos(\frac{\theta_k}{2})+j \sin(\frac{\theta_k}{2})
\end{eqnarray}
with real valued amplitude imbalance parameter $\lambda_k$ and phase imbalance parameter $\theta_k$. The signal is then input to a PA.

{The nonlinear distortion of PA can be 
characterized by the amplitude to amplitude conversion $A(|{x}^{a}_k(m)|)$ and amplitude to phase conversion $\phi(|{x}^{a}_k(m)|)$ \cite{ieee802}:
\begin{equation}
    A(|{x}^{a}_k(m)|)=  \frac{\alpha_a|{x}^{a}_k(m)|}{(1+(\alpha_a\frac{|{x}^{a}_k(m)|}{x_{\text{sat}}})^{2\sigma_a})^{\frac{1}{2\sigma_a}}},
\end{equation}
\begin{equation}
  \phi(|{x}^{a}_k(m)|)=\frac{\alpha_\phi |{x}^{a}_k(m)|^{q_1}}{1+ (\frac{|{x}^{a}_k(m)|}{\beta_{\phi}})^{q_2}},
\end{equation}
where  $\alpha_a$, $\alpha_{\phi}$, $\beta_{\phi}$, $\sigma_a$, $x_{\text{sat}}$, $q_1$ and $q_2$ are model parameters.} The distorted signal can then be expressed as 
\begin{equation} \label{newadd}
    s_k(m)=f({x}^{a}_k(m))=A(|{x}^{a}_k(m)|) e^{j(\text{angle}({x}^{a}_k(m))+\phi(|{x}^{a}_k(m)|))},
\end{equation}
{where $\text{angle}({x}^a_k)$ denotes the phase of the complex signal ${x}^a_k$. }

The received signal at time instant $m$ is represented as
\begin{equation}
    \mathbf{y}(m)=\mathbf{H}\mathbf{s}(m)+\boldsymbol{\omega}(m),\label{ym}
\end{equation}
where $\mathbf{H}\in \mathbb{C}^{N\times K}
$ is the MIMO channel matrix, $\mathbf{y}(m)=[y_1(m),y_2(m),\ldots,y_N(m)]^T$, $\mathbf{s}(m)= f({\mathbf{x}^a}(m))$ with ${\mathbf{x}^{a}}(m)= [{x}^{a}_1 (m),{x}^{a}_2 (m),\ldots,{x}^{a}_K (m)]^T$ being the length-$K$ vector, and $\boldsymbol{\omega}(m)$ denotes a white Gaussian noise vector. Note that the vectors and matrix in \eqref{ym} are all complex-valued, which can be rewritten as the following real model:
\begin{equation}
    \label{real_v}
    \underbrace {	\left[ \begin{matrix} \Re\{\mathbf{y}(m)\} \\ \Im\{\mathbf{y}(m) \} \end{matrix} \right]}_{\begin{huge}\mathbf{y}'(m)\end{huge}} =\! \underbrace {\left[ \begin{matrix} \Re\{\mathbf{H}\} & -\Im\{\mathbf{H} \}\\ \Im\{\mathbf{H} \} & \Re\{\mathbf{H}\}\end{matrix} \right]}_{\mathbf{H}'} \underbrace{\left[\begin{matrix} \Re\{\mathbf{s}(m)\} \\ \Im\{\mathbf{s}(m) \} \end{matrix}
    	\right]}_{\mathbf{s}'(m)} + \underbrace{\left[ \begin{matrix} \Re\{\boldsymbol{\omega}(m)\} \\ \Im\{\boldsymbol{\omega}(m) \} \end{matrix} \right]}_{\boldsymbol{\omega}'(m)}.  
\end{equation}
Due to the combined effects of I/Q imbalance and PA nonlinearity, the input-output relationship of the MIMO system is complex, and is denoted as
\begin{equation} \label{eqnew}
 \mathbf{y}'(m)=\mathcal{S}(\mathbf{x}(m))+\boldsymbol{\omega}'(m),  
\end{equation}
where $\mathcal{S}(\cdot)$ is the system transfer function. 

\rev{We assume that the channel matrix and the parameters of I/Q imbalance and PA nonlinearity models are unknown. Each user transmits a pilot signal followed by data. The aim of the receiver at the BS is to detect the transmitted data symbols of all users. To achieve this, there are two approaches.}
\begin{itemize} 
\item \rev{Direct detection: A symbol detector is trained directly using pilot symbols, where the input is the received signal and the output is the predicated symbols. As the system transfer function $\mathcal{S}(\cdot)$ is complicated, direct detection seems sensible. To deal with the nonlinearity, polynomial and DNN based techniques have been used in the literature. However, low order polynomials have limited capability to combat the nonlinearity. Although, high order polynomials have better capability, it is difficult to determine the polynomial coefficients due to numerical instability. The DNN techniques are more effective to deal with the nonlinearity, but it is difficult to train a DNN with a limited number of pilot symbols.}
\item \rev{Indirect detection: With the pilot symbols, the system function $\mathcal{S}(\cdot)$ is first identified, then a symbol detector is developed based on the system function. This strategy allows the design of powerful Bayesian detectors, but the implementation of indirect detection is challenging. First, to identify $\mathcal{S}(\cdot)$ with pilot symbols, we need to estimate the parameters of the I/Q imbalance and PA nonlinearity models and the MIMO channel at the same time, which is a difficult task due to the nonlinearity. Second, even if we assume that $\mathcal{S}(\cdot)$ is known, it is still difficult to develop a detector, especially a Bayesian one, due to the nonlinearity of $\mathcal{S}(\cdot)$. 
The aim of this work is to develop a Bayesian detector by using NN and factor graph techniques, which is more powerful than direct detection proposed in the literature.}
\end{itemize}


\subsection{Existing Detection Methods}
\subsubsection{Polynomial Based Direct Detection}
{A real-valued memory polynomial (RMP) model {was developed in} \cite{4773160},
where the I/Q branches after modulation are applied to the RMP model in order to compensate the I/Q imbalance. The work was extended to MIMO systems to address the joint effect of I/Q imbalance and PA nonlinearity in \cite{7600411}.}

{RMP can be used to directly compensate the hardware imperfections and deal with multi-user interference. The detector (for the $k$th user) can be expressed as
\begin{equation}
{\tilde x_k(m) = \mathrm{argmin}_{\lambda_a\in \mathcal{A}} |\hat{x}_{k}(m)-\lambda_a|}
\end{equation}
with
\begin{equation}\label{MP}
	\hat{x}_{k}(m) = \hat{x}_{k}^{\text{Q}}(m)+j \hat{x}_{k}^{\text{I}}(m)
\end{equation}
\begin{equation}\label{MP_q}
	\hat{x}_{k}^{\text{Q}}(m)\!=\!\sum_{n=1}^N\sum_{p=1}^P\sum_{l=0}^La_{p,l,k}^{\text{Q}} \Re\{y_n(m-l)\}^p+b_{p,l,k}^{\text{Q}} \Im\{y_n(m-l)\}^p
\end{equation}
\begin{equation}\label{MP_i}
	\hat{x}_{k}^{\text{I}}(m)\!=\!\sum_{n=1}^N\sum_{p=1}^P\sum_{l=0}^La_{p,l,k}^{\text{I}} \Re\{y_n(m-l)\}^p+b_{p,l,k}^{\text{I}} \Im\{y_n(m-l)\}^p,
\end{equation}
where $P$ is the order of the polynomial, {$L$ is the memory length,} and $\{a^{\text{Q}}_{p,l,k},a^{\text{I}}_{p,l,k}\}$ and $\{b^{\text{Q}}_{p,l,k}, b^{\text{I}}_{p,l,k}\}$ are the coefficients of the polynomial with respect to the real and imaginary parts of the received signals, respectively.} 

The RMP based detector is obtained by determining its polynomial coefficients $\{a^{\text{Q}}_{p,l,k},a^{\text{I}}_{p,l,k}\}$ and $\{b^{\text{Q}}_{p,l,k}, b^{\text{I}}_{p,l,k}\}$ using pilot signals. It is noted that models {\eqref{MP_q} and \eqref{MP_i} are} linear with respect to the polynomial coefficients. {With the mean squared error between $\{\hat{x}_{k}(m)\}$ and $\{{x}_{k}(m)\}$ as the cost function, the coefficients can be determined using least squares (LS)}. However, the determination of the coefficients suffers from numerical instability due to the involved matrix inversion, especially when the polynomial order is high \cite{1337325}.

\begin{figure}
	\begin{center}
		\includegraphics[width=3.5 in]{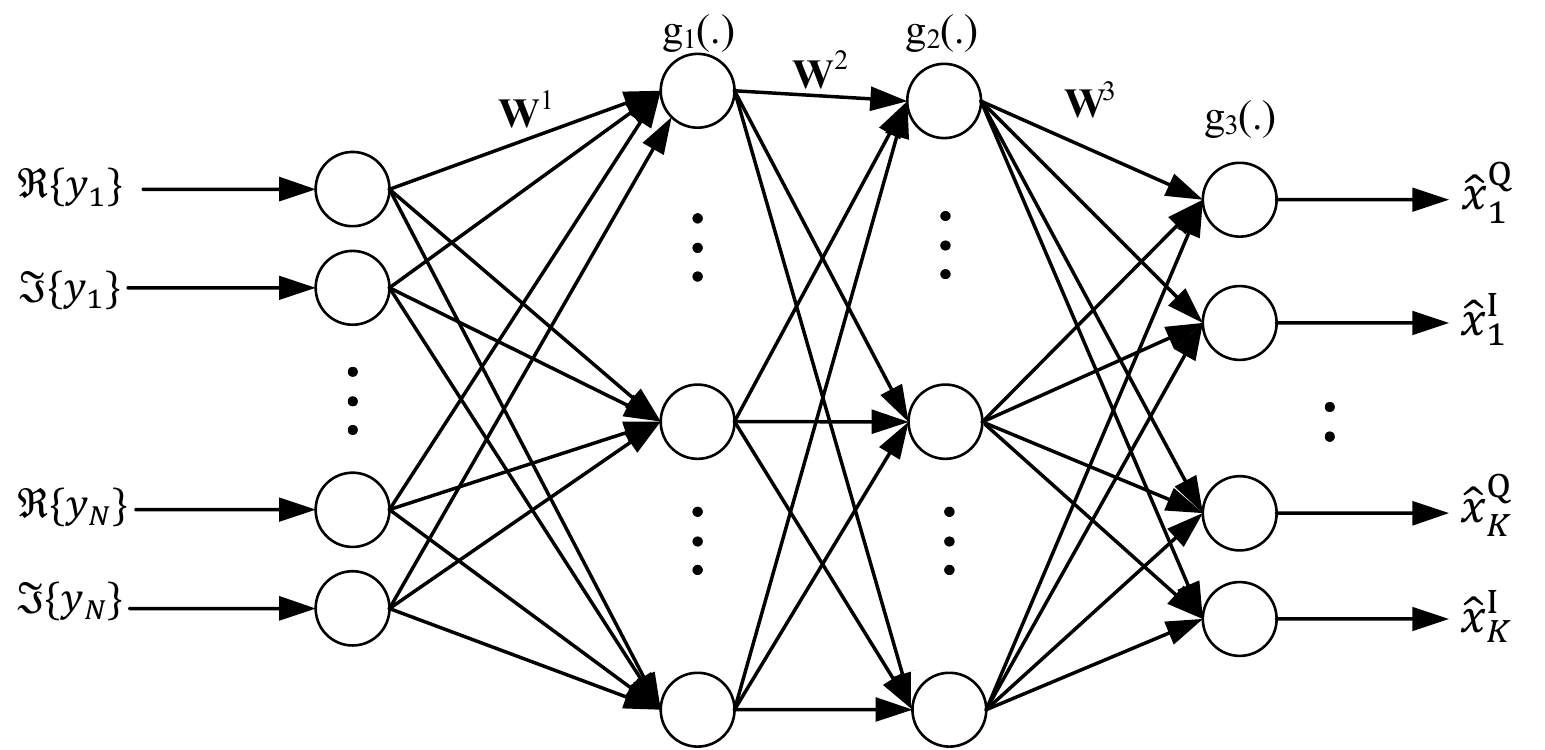}\\
		{\caption{{Illustration of DNN based direct detector.}\label{convDNN}}
		}
	\end{center}
\end{figure}

\subsubsection{DNN-Based Direct Detection}\label{Direct}
Another way to deal with the complex nonlinear relationship described in Section II.A is to use DNNs, leading to DNN-based detectors. 
As an example, a detector based on a real-valued DNN with two hidden layers is shown in Fig.~\ref{convDNN}, where the received signal is input to the DNN and estimated symbols are output, i.e.,
\begin{equation}
    \hat{\mathbf{x}}(m)=\mathcal{DNN}(\mathbf{y}'(m)),
\end{equation}
where the DNN deals with the combined distortions and multi-user interference. {A hard decision can be made based on $\hat{\mathbf{x}}(m)$, i.e., $\tilde x_k(m) = \mathrm{argmin}_{\lambda_a\in \mathcal{A}} |\hat{x}_{k}(m)-\lambda_a|$.}
Depending on the number of layers and hidden nodes, the number of parameters of the DNN can be large, leading to difficulties in training as a large number of pilot symbols are required. This results in an unacceptable overhead. The training of DNN receivers is prone to overfitting.   

\section{{Bayesian Signal Detection with Message Passing on Neural Networks}}

\rev{We adopt indirect detection and develop a Bayesian detector with the aid of NN and factor graph techniques. The development of the Bayesian detector relies on the signal model \eqref{eqnew}, in particular the system transfer function $\mathcal{S}(\cdot)$. However, it is difficult to estimate the unknown parameters and MIMO channel required in $\mathcal{S}(\cdot)$. To circumvent this, we train an NN (denoted by $\mathcal{NN}(\cdot)$ ) to substitute $\mathcal{S}(\cdot)$, i.e., we expect that
\begin{equation}\label{new1}
   \mathcal{NN}(\mathbf{x}) \approx \mathcal{S}(\mathbf{x}),
\end{equation}
for any symbol vector $\mathbf{x}$. The use of the substitute $\mathcal{NN}(\cdot)$ leads to the following benefits: 
\begin{itemize}
    \item Compared to estimating the parameters and MIMO channel involved in $\mathcal{S}(\cdot)$, the training of the NN is much easier, i.e., $\mathcal{NN}(\cdot)$ can be obtained using back-propagation. Moreover, the use of NNs is able to capture hardware imperfections that may not have explicit mathematical expressions. 
    \item Very different from the use of DNNs in the literature (which are typically a black box), the NN in this work is used to model $\mathcal{S}(\cdot)$. Hence, the architecture of the NN can be carefully designed based on the signal flow of the MIMO system as detailed in Section III.A, so that the number of parameters of the NN can be minimized, which is crucial to achieving efficient training with limited number of pilot symbols.
    \item Bayesian inference based on $\mathcal{NN}(\cdot)$ is easier than that based on $\mathcal{S}(\cdot)$ as the buliding blocks of $\mathcal{NN}(\cdot)$ are matrix-vector products and activation functions. We will show in Section III.B that, leveraging UAMP, efficient Bayesian inference for symbol detection can be implemented with message passing.  
\end{itemize}
} 

\subsection{{Signal Flow Based NN Architecture Design and Training}}
 \begin{figure}
 	\begin{center}
    \includegraphics[width=3.5 in]{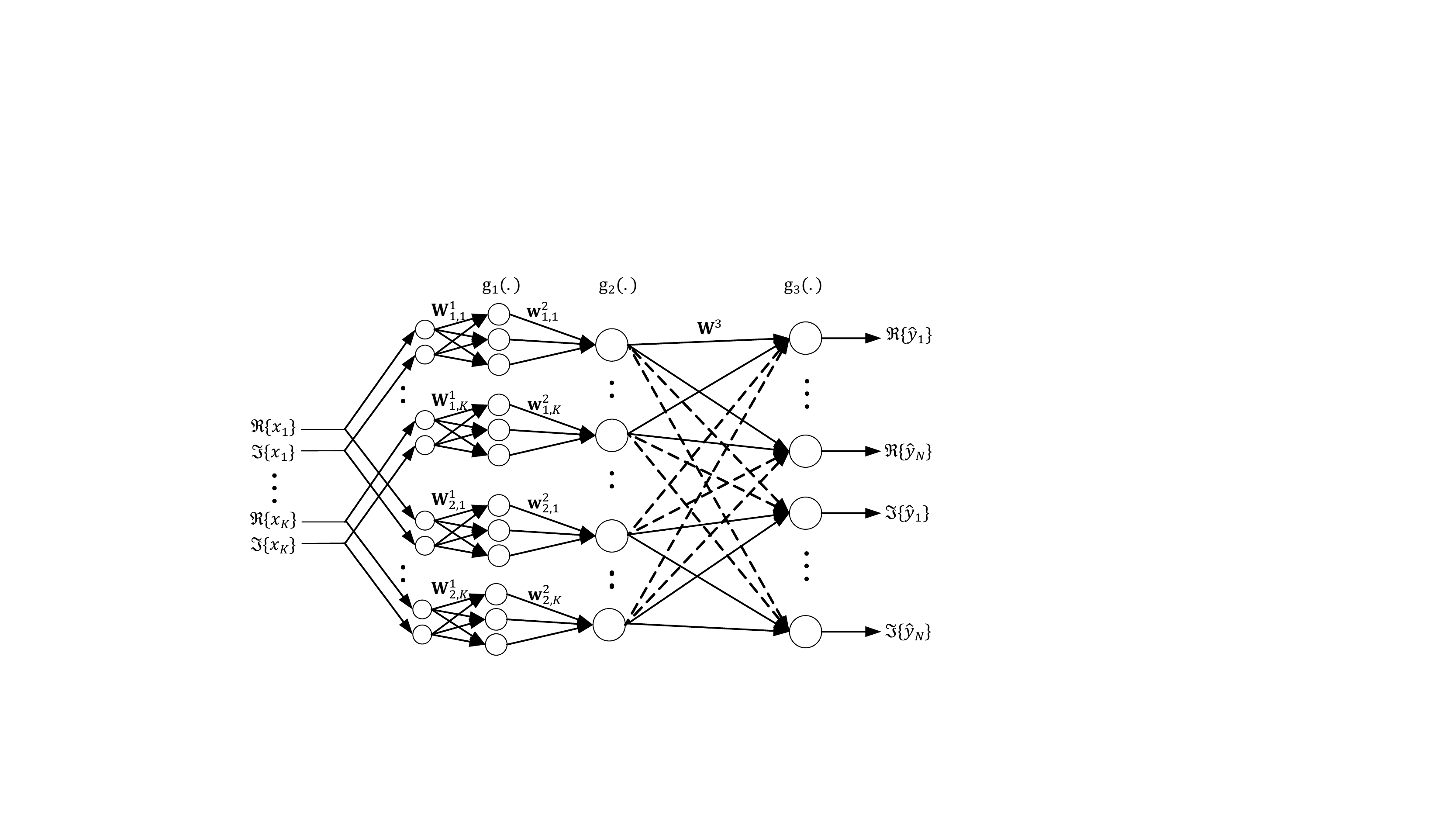}\\
 		\caption{Proposed NN to characterize hardware imperfections and multi-user interference. }\label{nn}
 	\end{center}
 \end{figure}
 
As shown in Fig. \ref{nn}, the NN consists of an input layer, two non-fully connected hidden layer and an output layer. {We note that the NN is used to model the system characterized by \eqref{newimbalance}, \eqref{newadd} and \eqref{real_v}. The symbols of all users are input to the NN, and the outputs of the NN are the predicted received signals, where the real and imaginary parts of the signals are separated to make the NN a real-valued one. The architecture of the NN is designed based on the signal flow expressed with \eqref{newimbalance}, \eqref{newadd} and \eqref{real_v}, i.e., the transmitted symbols are first distorted due to I/Q imbalance and PA nonlinearity and then undergo multi-user interference.} 

In Fig. \ref{nn}, the input layer, the first hidden layer and the input to the second hidden layer are essentially $2K$ sub-NNs, which are used to model the I/Q imbalance and PA nonlinearity of the $K$  users. Each sub-NN has two input nodes corresponding to the real and imaginary parts of a symbol, and a single hidden layer with $N'$ hidden nodes, where the activation function \textit{Tanh} is employed. As shown in Fig. \ref{nn}, the real and imaginary parts of a symbol are shared by two sub-NNs, which are called a sub-NN pair. There are $K$ sub-NN pairs in total, {and they are indexed by $(l,k)$, where $l=1,2$ and $k=1, ..., K$. The pair of sub-NN $(1,k)$ and sub-NN $(2,k)$} models the combined I/Q imbalance and PA nonlinearity of user $k$ shown in \eqref{newimbalance} and \eqref{newadd}, i.e., the output of one sub-NN is expected to be a good approximation to $\Re\{s_k(m)\}$ and the output of the other one is expected to be a good approximation to $\Im\{s_k(m)\}$. More details are explained in the following. 

         

{According to Fig. \ref{nn},  
the input to the $k$th sub-NN pair is denoted as
\begin{equation}
 \mathbf{c}_{k}(m)=[\Re\{{x}_k(m)\},\Im\{{x}_k(m)\}]^T.
\end{equation}
Then, the output of the $(l,k)$th sub-NN is 
\begin{equation}
 \mathbf{d}_{l,k}(m)={g_1}({\mathbf{W}^{1}_{l,k}}\mathbf{c}_k(m) + {\mathbf{b}^{1}_{l,k}}),
\end{equation}
where $\mathbf{W}^{1}_{l,k}$ and $\mathbf{b}^{1}_{l,k}$ are the corresponding weight matrix and bias vector of the sub-NN $(l,k)$, and $\mathbf{W}^{1}_{l,k}=[\mathbf{w}^{1}_{l,k,1},\mathbf{w}^{1}_{l,k,2}]^T$ 
with $\mathbf{w}^{1}_{l,k,1}=[{w}^{1}_{l,k,11},{w}^{1}_{l,k,12},\ldots,{w}^{1}_{l,k,1N'}]^T$ and $\mathbf{w}^{1}_{l,k,2}=[{w}^{1}_{l,k,21},{w}^{1}_{l,k,22},\ldots,{w}^{1}_{l,k,2N'}]^T$. 
Each sub-NN has one output node, 
and the output of the $(l,k)$th sub-NN can be expressed as 
\begin{equation}
s_{l,k}(m) = {({\mathbf{w}^{2}_{l,k}})^T}\mathbf{d}_{l,k}(m),
\end{equation}
where $\mathbf{w}^{2}_{l,k}=[{w}^{2}_{l,k,1},{w}^{2}_{l,k,2},\ldots,{w}^{2}_{l,k,N'}]^T$  are the output weights of a sub-NN.}
{It is known that an NN with a single hidden layer has {the property of universal approximation \cite{huang1998upper}}. We find that the sub-NNs with a single hidden layer in Fig. \ref{nn} are sufficient to model the combined PA nonlinear distortion and I/Q imbalance.} 
{It is noted that, when all transmitters have the same I/Q imbalance and PA nonlinearity, the sub-NN pairs share the weight and bias parameters, i.e., the parameters of the sub-NN pairs can be tied. This reduces the number of parameters of all sub-NNs from $6KN'$ to $6N'$.}


{
Assume that the combined I/Q imbalance and PA nonlinearity are well modelled using the sub-NNs. The second hidden layer and the output layer are designed to model the multi-user interference. 
The activation functions of the two layers $g_2(.)$ and $g_3(.)$ are linear as the interference shown in \eqref{real_v} is in a linear form. The second hidden layer is fully connected to the output layer, yielding the predicted in-phase and quadrature components of the received signal.} 
{Considering the structure of $\mathbf{H}'$ in (\ref{real_v}), the weight matrix $\mathbf{W}^3$ between the second hidden layer and output layer should have the same structure. To impose such a structure on the weight matrix, we can tie the elements of the weight matrix properly, leading to the following weight matrix:} 
\begin{equation}\label{cw1}
\mathbf{W}^{3}=
\begin{bmatrix}
 \mathbf{W}^{31}& \mathbf{W}^{32}  \\
 -\mathbf{W}^{32} &  \mathbf{W}^{31} \end{bmatrix},
\end{equation}
{where $\mathbf{W}^{31}$ and $\mathbf{W}^{32}$ are sub-weight matrices with dimension $N \times K$. It can be seen that the weight matrix has $2KN$ parameters, which is in contrast to the unstructured weight matrix that has $4KN$ parameters. 
{Then the output of the NN can be expressed as  
  \begin{equation}\label{denseconnect}
   \hat{\mathbf{y}}'(m)=\mathbf{W}^3\mathbf{s}'(m),
\end{equation} 
where $\mathbf{s}'(m)=[s_{1,1}(m),\ldots,s_{1,K}(m),\ldots,s_{2,K}(m)]^T$ is the output vector from the $2K$ sub-NNs, and $\hat{\mathbf{y}}'(m)=[v_{1,1}(m),\ldots,v_{1,N}(m),\ldots,v_{2,N}(m)]^T$ is a length-$2N$ output vector with ${v}_{1,n}(m)=\Re\{\hat{y}_{n}(m)\}$ and ${v}_{2,n}(m)=\Im\{\hat{y}_{n}(m)\}$.
Hence, the predicted signal of the $n$th receive antenna is represented as $\hat{y}_{n}(m) = {v}_{1,n}(m)+j {v}_{2,n}(m)$.
}

{The training of the NN is straightforward. Suppose that the length of the pilot signal is $M_0$, i.e., we have $M_0$ training samples $\{(\mathbf{p}(m),\mathbf{t}(m)), m=1,\ldots,M_0\}$,
where  $\mathbf{t}(m)=[{t}_{1}(m),{t}_{2}(m),\ldots,{t}_{K}(m)]^T$ and $\mathbf{p}(m)=[{p}_{1}(m),{p}_{2}(m),\ldots,{p}_{N}(m)]^T$ denote the pilot symbols and corresponding received signal. With the input $\mathbf{t}'(m)=[\Re\{\mathbf{t}(m)\}^T,\Im\{\mathbf{t}(m)\}^T]^T$, the expected output $\mathbf{p}'(m)=[\Re\{\mathbf{p}(m)\}^T,\Im\{\mathbf{p}(m)\}^T]^T$ and loss function} 
\begin{equation}
    \text{Loss}=\frac{1}{2N}\frac{1}{M_0}\sum^{M_0}_{m=1}\sum^{2N}_{n=1}(v_n(m)-p'_n(m))^2,
\end{equation}
{the NN can be trained, i.e., the weights $\{\mathbf{W}^{1}_{l,k},\mathbf{w}^{2}_{l,k},\mathbf{W}^{3}\}$ and biases $\{\mathbf{b}^{1}_{l,k}\}$ are determined with back-propagation \cite{58323}.} 

After training, we obtain the following model:
\begin{eqnarray}\label{dnnmodel}
    \mathbf{y}'(m)&=&\hat{\mathbf{y}}'(m)+\boldsymbol{\omega}'(m) \nonumber  \\ 
    &=&\mathcal{NN}(\mathbf{x}(m))+\boldsymbol{\omega}'(m),
\end{eqnarray}
where $\mathcal{NN}(\cdot)$ denotes the trained NN and the term $\boldsymbol{\omega}'(m)$ denotes a noise vector that also accounts for training and modelling errors. Then we are ready to detect the transmitted symbols based on the trained NN, which is elaborated in the next section. 

\subsection{{Bayesian Signal Detection Based on the Trained NN}}

\begin{algorithm}\caption{MP-NN Message Passing Detector}\label{Alg}
Define vector $\boldsymbol{\lambda} = \mathbf{\Lambda \Lambda}^H \textbf{1}$. Initialization: $\tau_s^{(0)} = 1$, $\hat{\mathbf{s}}^{(0)} = \textbf{0}$, $\mathbf{c} = \textbf{0}$, $\hat{\mathbf{x}}^{(0)}=\textbf{0}$ , $\hat{\epsilon}=1$ and $i = 0$.\\
\textbf{Repeat}
\begin{algorithmic}[1]

    
    

\STATE      $\boldsymbol{\tau}_{p} = \tau_{s}^{i} \boldsymbol{\lambda}$\;\label{LINE5}
      
    \STATE     ${\mathbf{p}} = \mathbf{\Phi} {\hat{\mathbf{s}}}^{i} - \boldsymbol{\tau}_{p} \cdot \mathbf{c}$\;\label{LINE6}
      
        \STATE $\mathbf{\tau}_z=\boldsymbol{\tau}_p./(\mathbf{1}+\hat{\epsilon}\boldsymbol{\tau}_p)$\;\label{LINE7}
          \STATE
          $\hat{\mathbf{z}}=(\hat{\epsilon}\boldsymbol{\tau}_p \cdot\mathbf{r} +\mathbf{p})./(\boldsymbol{1}+\hat{\epsilon}\boldsymbol{\tau}_p)$\; \label{LINE8}
          \STATE
$\hat{\epsilon}={2N}/(||\mathbf{r}-\hat{\mathbf{z}}||^2+\boldsymbol{1}^{H}\mathbf{v}_{z})$\; \label{LINE9}

 \STATE        ${\boldsymbol{\tau}_c} = \boldsymbol{1}./(\boldsymbol{\tau}_p+\hat{\epsilon}^{-1}\boldsymbol{1})$\;\label{LINE10}
 \STATE        ${\mathbf{c}} = \boldsymbol{\tau}_{c} \cdot ( \mathbf{r}-\mathbf{p})$\;\label{LINE11}

 \STATE        $1/\tau_{q} = (1/2K)  \boldsymbol{\lambda}^H \boldsymbol{\tau}_{c}$\;\label{LINE12}
      
 \STATE        ${{\mathbf{q}}} = \hat{\mathbf{s}}^i + \tau_{q} (\mathbf{\Phi}^H \mathbf{c})$\;\label{LINE13}
        \STATE $\forall l, k$, $\tilde{q}^{i}_{l,k}=(\mathbf{w}^{2}_{l,k})^Tg(\mathbf{W}^{1}_{l,k}\hat{\mathbf{x}}'_k+\mathbf{b}^{1}_{l,k})$\;\label{LINE16}
     \STATE  $\forall l, k$, $\eta^{i}_{l,k}=(\mathbf{w}^{2}_{l,k}\cdot\mathbf{w}^{1}_{l,k,1})^Tg'(\mathbf{W}^{1}_{l,k}\hat{\mathbf{x}}'_k+\mathbf{b}^{1}_{l,k})$,\;\label{LINE14}
              \STATE $\forall l, k$, $\gamma^{i}_{l,k}=(\mathbf{w}^{2}_{l,k}\cdot\mathbf{w}^{1}_{l,k,2})^Tg'(\mathbf{W}^{1}_{l,k}\hat{\mathbf{x}}'_k+\mathbf{b}^{1}_{l,k})$\;\label{LINE15}


 \STATE         $\forall l, k$,   
 ${\tau}^{l,1}_{{\psi}_{l,k}} = ({\tau}_{q}+\gamma_{l,k}^2{\tau}_{{x}_{l'(l'\ne l),k}})/{\eta}^2_{l,k}$\;\label{LINE17}
  \STATE   $\forall l, k$,
 ${\tau}^{l,2}_{{\psi}_{l,k}} = ({\tau}_{q}+\eta_{l,k}^2{\tau}_{{x}_{l'(l'\ne l),k}})/{\gamma}^2_{l,k}$\;\label{LINE18}
 
    \STATE   $\forall l, k$,      ${\psi}^{l,1}_{l,k} =({q}_{l,k}-\tilde{q}_{l,k})/\eta_{l,k} +\hat{x}_{l,k}$\;\label{LINE19}
        \STATE      $\forall l, k$,   ${\psi}^{l,2}_{l,k} =({q}_{l,k}-\tilde{q}_{l,k})/\gamma_{l,k} +\hat{x}_{l,k}$\;\label{LINE20}

    \STATE   $\forall l, k$,    ${\tau}_{{\psi}_{l,k}} =( 1/{{\tau}_{\psi_{l,k}}^{l,1}}+1/{{\tau}_{\psi_{l,k}}^{l,2}})^{-1}$\;\label{LINE21}

        \STATE     $\forall l, k$,     ${\psi}_{l,k} =({{\psi}_{l,k}^{l,1}}/{{\tau}_{\psi_{l,k}}^{l,1}}+{{\psi}_{l,k}^{l,2}}/{{\tau}_{\psi_{l,k}}^{l,2}}){\tau}_{{\psi}_{l,k}}$\;\label{LINE22}

        \STATE     $\forall  k$,     ${\tau}_{\tilde{\psi}_{k}} ={\tau}_{{\psi}_{1,k}}+{\tau}_{{\psi}_{2,k}}$\;\label{LINE22_2}
        \STATE     $\forall  k$,     $\tilde{\psi}_{k} ={\psi}_{1,k}+j{\psi}_{2,k}$\;\label{LINE22_1}
  \STATE   $\forall k,a$,    $\xi_{k,a} = \text{exp}(-{\tau}_{\tilde{\psi}_{k} }^{-1}|\lambda_a-\tilde{\psi}_{k}|^2)$\;\label{LINE23}
 \STATE     $\forall k, a$,          $\mu_{k,a} = \xi_{k,a}/\sum^{|A|}_{a=1}\xi_{k,a} $\;\label{LINE24}
 \STATE      $\forall k$,     $\hat{x}^{i+1}_{k} =\sum^{|A|}_{a=1}\lambda_{a}\mu_{k,a} $\;\label{LINE25}
 
  \STATE     $\forall k$,  $\tau_{x^{i+1}_{k}} =\sum^{|A|}_{a=1}\mu_{k,a}|\lambda_a-\hat{x}^{i+1}_k|^2 $\;\label{LINE26}
  
    \STATE  Calculate $\eta^{i+1}_{l,k}, \gamma^{i+1}_{l,k}$, $\tilde{q}^{i+1}_{l,k}$ again using Lines 10-12 with $\hat{x}^{i+1}_{k}$\;\label{LINE27}

 \STATE  $\forall k$, 
 ${\tau}^{i+1}_{{x}_{1,k}} ={\tau}^{i+1}_{{x}_{2,k}}=1/2\tau^{i+1}_{x_{k}}$\;\label{LINE27_1}
  \STATE  $\forall k$, 
 $\hat{x}^{i+1}_{1,k}=\Re\{\hat{x}^{i+1}_{k}\},\hat{x}^{i+1}_{2,k}=\Im\{\hat{x}^{i+1}_{k}\} $\;\label{LINE27_2}
   \STATE  $\forall l, k$, 
 ${\tau}^{i+1}_{s_{l,k}} =(\eta^{i+1}_{l,k})^2{\tau}^{i+1}_{{x}_{1,k}}+(\gamma^{i+1}_{l,k})^2{\tau}^{i+1}_{{x}_{2,k}}$\;\label{LINE28}
    \STATE      $\forall l, k$,   $\hat{s}^{i+1}_{l,k}=\tilde{q}^{i+1}_{l,k}$\;\label{LINE28_1}
  
\STATE $\tau_s^{i+1} = \frac{1}{4K}\sum_{l=1}^{2}\sum_{k=1}^{2K}{\tau}^{i+1}_{s_{l,k}} $\;\label{LINE29}
 \STATE     $i = i+1$\;\label{LINE30}
\end{algorithmic}
\textbf{Until terminated}
\end{algorithm}

{During the phase of data transmission, we perform Bayesian inference for the transmitted symbols based on the {trained} NN, i.e., model \eqref{dnnmodel}. It is noted that the error term $\boldsymbol{\omega}'(m)$ is unknown. To deal with this, we assume that it is white Gaussian with mean zero and unknown variance $\epsilon^{-1}$ ($\epsilon$ is called precision). 
Our aim is to determine the transmitted symbol vector $\mathbf{x}(m)$ based on the received signal $\mathbf{y}(m)$. We use the Bayesian approach, in particular, the message passing techniques, where we represent the trained NN \eqref{dnnmodel} as a factor graph. The weight matrix {$\mathbf{W}^3$} in the NN leads to a densely connected factor graph, resulting in difficulties in message passing in terms of complexity and convergence. The AMP algorithm is efficient in handling short loops induced by i.i.d. (sub-)Gaussian matrices, but the weight matrix $\mathbf{W}^3$ here is not i.i.d. (sub-)Gaussian, making the AMP algorithm easily diverge. Therefore, we use the UAMP algorithm.}

{According to \eqref{denseconnect} and \eqref{dnnmodel}, we have
\begin{equation} \label{newmodel}
    \mathbf{y}'=\mathbf{W}^3\mathbf{s}' +\boldsymbol{\omega}',    
\end{equation}
where the time index $m$ is dropped for the simplicity of notation. 
As UAMP works with a unitary transformed model, we perform a unitary transformation to (\ref{newmodel}), i.e., } 
{
\begin{equation}
	\mathbf{r} = \mathbf{U}^H\mathbf{y}' =\mathbf{\Phi}\mathbf{s}' + \tilde{\boldsymbol{\omega}},
\end{equation}
where $\mathbf{\Phi}=\mathbf{U}^H{\mathbf{W}^3}= \mathbf{\Lambda V }$,  $\mathbf{U}$ is obtained from the SVD ${\mathbf{W}^3} = \mathbf{U \Lambda V}$, 
 and the noise $\tilde{\boldsymbol{\omega}}=\mathbf{U}^H{\boldsymbol{\omega}}'$ has the same distribution as  $\boldsymbol{\omega}'$ since $\mathbf{U}$ is an unitary matrix. The precision of the noise is still denoted by $\epsilon$. As the noise precision $\epsilon$ is unknown, its estimation is included in the detector. }
Define an auxiliary vector
\begin{equation}
\mathbf{z}=\mathbf{\Phi}\mathbf{s}', 
\end{equation}
which is treated as a latent variable.  
{Then the joint distribution of $\mathbf{x}$, $\mathbf{s}'$, $\mathbf{z}$ and $\epsilon$ given $\mathbf{r}$ can be expressed as}
\begin{equation}\label{fac}
\begin{aligned}
&p(\mathbf{x},\mathbf{z},\mathbf{s}', \epsilon|\mathbf{r})
\propto p(\epsilon) p(\mathbf{r}|\mathbf{z},\epsilon) p(\mathbf{z}|\mathbf{s}') p(\mathbf{s}'|\mathbf{x}){p(\mathbf{x})},   \\
\end{aligned}
\end{equation}
where we assume an improper prior for the noise precision, i.e., $p(\epsilon) \propto 1/\epsilon$ {\cite{tipping2001sparse}},
\begin{equation}
   p(\mathbf{r}|\mathbf{z},\epsilon)=\prod_n p(r_n|z_n,\epsilon)  
\end{equation}
with $p(r_n|z_n,\epsilon)=\mathcal{N}(z_n;r_n,\epsilon^{-1})$, 
\begin{equation}
p(\mathbf{z}|\mathbf{s}')=\delta(\mathbf{z}-\mathbf{\Phi}\mathbf{s}')=\prod_n \delta(z_n-\mathbf{\Phi}^T_n\mathbf{s}'), 
\end{equation}
with $\mathbf{\Phi}^T_n$ being the $n$th row of $\mathbf{\Phi}$,
\begin{equation}
    p(\mathbf{s}'|\mathbf{x})=\prod_{l,k}p({s}_{l,k}|{x}_{1,k},{x}_{2,k})={\prod_{l,k}\delta({s_{l,k}-f_l(\mathbf{x}'_k)})},
\end{equation}
{with $f_l(\mathbf{x}'_k)$ given later in \eqref{flxkp} and $\mathbf{x}'_k=[x_{1,k},x_{2,k}]^T$},
and
\begin{equation}\label{newx}
    {p(\mathbf{x})}=\prod_kp(x_k)={\prod_k(1/|\mathcal{A}|)\sum^{|\mathcal{A}|}_{a=1}\delta(x_{k }-\lambda_a)}.
\end{equation}
Our aim is to obtain the (approximate) marginal (a posteriori distribution) of each transmitted symbol $p(x|\mathbf{r})$, based on which a hard decision can be made with the maximum posterior probability (MAP) criterion.

\begin{figure}
	\begin{center}
		\includegraphics[width=3.5 in]{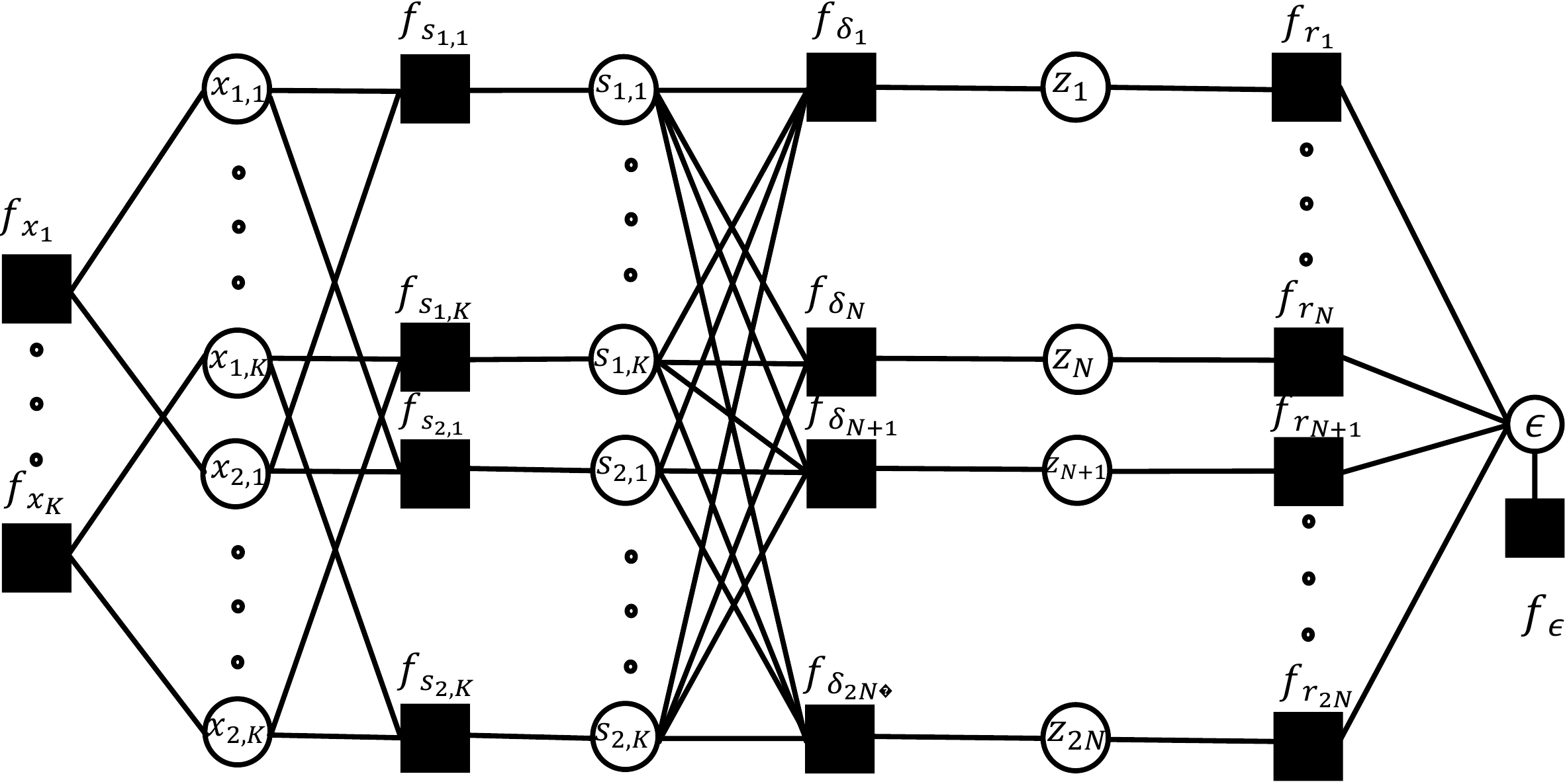}\\
		\caption{{ {Factor graph representation of the NN-modeled system.}}}\label{mimo}
	\end{center}
\end{figure}

The factor graph representation for the factorization in (\ref{fac})-\eqref{newx} is depicted in Fig. \ref{mimo}, where squares and circles represent function nodes and variable nodes, respectively. To facilitate the factor graph
representation, we introduce the
notations in Table \ref{tab:table1}, which shows the correspondence between the
factor labels and the corresponding distributions they represent.

\begin{table}[h!]
  \begin{center}
    \caption{Factors, Underlying Distributions and Functional Forms Associated with (\ref{fac})}
    \label{tab:table1}
    \begin{tabular}{l|c|r} 
      \textbf{Factor} & \textbf{Distribution} & \textbf{Functional Form}\\
      \hline $f_{r_n}$ & $p(r_n|z_n,\epsilon)$ & $\mathcal{N}(z_n;r_n,\epsilon^{-1})$\\            $ f_{\delta_n} $ & $p(z_n|\mathbf{s})$   &  $\delta(z_n-\mathbf{\Phi}_n\mathbf{s}')$\\
        $ f_{s_{l,k}} $ & $p({s}_{l,k}|{x}_{1,k},{x}_{2,k})$   &  $\delta({s}_{l,k}-f_l(\mathbf{{x}'_k}))$\\
      $f_{x_{k}}$ & $p(x_{k})$ & $(1/|\mathcal{A})|\sum^{|\mathcal{A}|}_{a=1}\delta(x_{k }-\lambda_a)$\\

      $f_\epsilon$ & $p(\epsilon)$ & $\propto \epsilon^{-1}$\\
      \hline
      
    \end{tabular}
  \end{center}
\end{table}

 We develop a message passing algorithm based on the factor graph in Fig. \ref{mimo}. Due to the presence of loops in the graph, an iterative process is required, which involves several rounds of forward and backward recursions. In particular, we use UAMP to handle the densely connected part of the graph, which is crucial to achieving high performance while with low complexity. To deal with various factor nodes, both belief propagation (BP) \cite{2009pans} and variational message passing (VMP) \cite{winn2005variational} are used. In the following we derive the message updates, where the message passed from node $A$ to node $B$ is denoted by $m_{A\rightarrow B}(c)$, which is a function of $c$. It is noted that the message passing algorithm is an iterative one, and some message computations in the current iteration require messages computed in the last iteration. The message passing algorithm is summarized in Algorithm \ref{Alg}, and the derivations of the algorithm line by line are elaborated in the following.      
{According to the derivation of (U)AMP using loopy BP, (U)AMP provides the message from variable node $z_m$ to function node $f_{r_m}$. Due to the Gaussian approximation in the the
derivation of (U)AMP, the message is Gaussian, i.e.,
\begin{equation}
m_{z_{n}\rightarrow f_{r_n}}(z_n)= m_{f_{\delta_n}\rightarrow z_n}(z_n) \propto  \mathcal{N}(z_{n};{p}_n,\tau_{p_{n}}),
\end{equation}
where the mean $p_n$ and the variance $\tau_{p_n}$ are the $n$th elements of $\mathbf{p}$ and $\boldsymbol{\tau}_p$ given in Lines \ref{LINE5} and \ref{LINE6} of Algorithm 1. 
}

{Following VMP, the message $m_{f_{r_n}\rightarrow \epsilon}(\epsilon)$ from factor
node $f_{r_n}$ to variable node $\epsilon$ can be expressed as \begin{equation}\label{frnep}
m_{f_{r_{n}}\rightarrow \epsilon}(\epsilon) \propto \exp \left\{\langle \log f_{r_{n}}\left(z_{n}, \epsilon\right) \rangle_{b(z_n)} \right\},
\end{equation}
where the belief of $z_n$ is given as \begin{equation}
b(z_{n}) \propto  m_{{z_n}\rightarrow f_{r_n}}(z_n)  m_{f_{r_n}\rightarrow {z_n}}(z_n).
\end{equation}
Later in \eqref{rznrreal}, we will show that $m_{f_{r_n}\rightarrow {z_n}}(z_n) \propto \mathcal{N}(z_n;r_n,\hat{\epsilon}^{-1})$ with $\hat{\epsilon}^{-1}$ being the estimate of $\epsilon^{-1}$ in last iteration, and its computation is given in \eqref{epphat}. Hence $b(z_n)$ is Gaussian according to the property of the product of Gaussian functions, i.e., $b(z_{n}) =\mathcal{N}(z_n;\hat{z}_n,v_{z_n})$ with 
\begin{equation}\label{vzn}
v_{z_n} = (1/\tau_{p_n}+\hat{\epsilon})^{-1}
\end{equation}
\begin{equation}\label{hzm}
\hat{z}_n = v_{z_n}(\hat{\epsilon}r_n+p_n/\tau_{p_n}).
\end{equation}
Note that $\boldsymbol{\tau}_p$ may contain zero elements. To avoid numerical problems in \eqref{vzn} and \eqref{hzm}, they can be rewritten (in vector form) as
\begin{equation}\label{vznv}
\boldsymbol{\tau}_z=\boldsymbol{\tau}_p./(\mathbf{1}+\hat{\epsilon}\boldsymbol{\tau}_p),
\end{equation}
\begin{equation}\label{hzmv}
\hat{\mathbf{z}}=(\hat{\epsilon}\boldsymbol{\tau}_p \cdot\mathbf{r} +\mathbf{p})./(\boldsymbol{1}+\hat{\epsilon}\boldsymbol{\tau}_p),
\end{equation}
which are Lines 3 and 4 of Algorithm 1.
From \eqref{frnep} and the Gaussianity of $b(z_n)$, the message $m_{f_{r_n}\rightarrow \epsilon}(\epsilon)$ can be expressed as 
\begin{equation}\label{frmbeta}
m_{f_{r_{n}}\rightarrow \epsilon}(\epsilon) 
\propto \sqrt{\epsilon} {\exp}(-\frac{\epsilon}{2}(|r_n-\hat{z}_n|^2+v_{z_n})).
\end{equation}
According to VMP, the message from function node $f_{r_n}$ to variable node $z_n$ is 
\begin{equation}\label{rznrreal}
\begin{aligned}
m_{f_{r_{n}}\rightarrow z_{n}}(z_{n}) &\propto  \exp \left\{\langle \log f_{r_{n}}\left(z_{n}, \epsilon\right) \rangle_{b(\epsilon)} \right\} \\
&\propto  \mathcal{N}\left(z_{n} ; r_{n}, \hat{\epsilon}^{-1}\right),
\end{aligned}
\end{equation}
where $\hat{\epsilon}=\langle\epsilon\rangle_{b(\epsilon)}$ with
\begin{equation}\label{bep}
	\begin{aligned}
		b(\epsilon) &\propto  m_{\epsilon\rightarrow f_{r_n}}(\epsilon)m_{f_{r_n}\rightarrow\epsilon}(\epsilon)\\
		&= f_{\epsilon}(\epsilon) \prod_n^{2N}m_{f_{r_{n}}\rightarrow \epsilon}(\epsilon)\\
		&\propto \epsilon^{N-1} \exp\left\{-\frac{\epsilon}{2}\sum_n\left(|r_n-\hat{z}_n|^2+v_{z_n}\right)\right\}
	\end{aligned}
\end{equation}
and 
\begin{equation}\label{frmbeta_1}
m_{\epsilon\rightarrow f_{r_n}}(\epsilon)=f_{\epsilon}(\epsilon)\prod_{n'\neq n} m_{f_{r_{n'}}\rightarrow \epsilon}(\epsilon).
\end{equation}
It is noted that $b(\epsilon)$ follows a Gamma distribution with rate parameter $-\frac{1}{2}\sum_n\left(|r_n-\hat{z}_n|^2+v_{z_n}\right)$ and shape parameter $N$, so $\hat{\epsilon}=\langle\epsilon\rangle_{b(\epsilon)}$ can be computed as 
\begin{equation}\label{epphat}
\hat{\epsilon}=\frac{2N}{\sum_{n=1}^{2N}(|r_n-\hat{z}_n|^2+v_{z_n})},
\end{equation}
which can be rewritten in vector form shown in Line 5 of Algorithm 1. From \eqref{rznrreal}, the Gaussian form of the message $m_{f_{r_{n}}\rightarrow z_{n}}(z_{n})$ suggests the following model
	\begin{equation}\label{betahat}
		r_n=z_n+\omega_n, n=1,\ldots, 2N,
	\end{equation}
where $w_n$ is a Gaussian noise with mean 0 and variance $\hat{\epsilon}^{-1}$. This fits the forward recursion of the UAMP algorithm with
a known noise variance, corresponding to Lines 6 - 9 of Algorithm 1. }

{According to the derivation of UAMP, it produces the message $m_{s_{l,k}\rightarrow f_{s_{l,k}}}(s_{l,k})\propto \mathcal{N}(s_{l,k};q_{l,k},\tau_{q})$ with mean  $q_{l,k}$ and variance $\tau_q$, which are given in Lines 8 and 9 of Algorithm 1.}
{Next, we need to compute the outgoing message of the function node $f_{s_{l,k}}=\delta({s}_{l,k}-f_l(\mathbf{{x}'_k}))$. It is noted that the local function is nonlinear with the following expression:  
	\begin{equation}\label{flxkp}
		{f}_l(\mathbf{x}'_k) =(\mathbf{w}^{2}_{l,k})^Tg_1( {{\mathbf{w}^{1}_{l,k,1}}x_{1,k} + {\mathbf{w}^{1}_{l,k,2}}x_{2,k} + {\mathbf{b}^{1}_{l,k}}}),
	\end{equation}
	where $g_1(\cdot)=Tanh(\cdot)$. The nonlinear function makes the computation of the message $m_{f_{s_{l,k}}\rightarrow x_{l,k}}(x_{l,k})$ intractable. To solve this problem, ${f}_l({\mathbf{x}'_k})$ is linearized by using the first order Taylor expansion at the estimate of $\mathbf{x}'_k$ in the last iteration, i.e.,
	\begin{equation}\label{flxk}
		{f}_l({\mathbf{x}'_k}) \approx {f}_l(\hat{\mathbf{x}}'_k) + {f}_l^{'}(\hat{\mathbf{x}}'_k)(\mathbf{x}'_k- \hat{\mathbf{x}}'_k)
	\end{equation}
	with 
	\begin{equation}\label{qlk}
		{f}_l(\hat{\mathbf{x}}'_k)=\tilde{q}_{l,k} =(\mathbf{w}^{2}_{l,k})^Tg_1(\mathbf{W}^{1}_{l,k}\hat{\mathbf{x}}'_k+\mathbf{b}^{1}_{l,k}),
	\end{equation} 
	which is Line 10 of Algorithm 1, and
	\begin{equation}\label{5.30}
		{f}_l^{'}(\hat{\mathbf{x}}'_k) = {\left[ {\frac{{\partial {{f}_l(\hat{\mathbf{x}}'_k)}}}{{\partial x_{1,k }}},\frac{{\partial {{f}_l(\hat{\mathbf{x}}'_k)}}}{{\partial x_{2,k}}}} \right]^T} = {\left[ {\eta_{l,k} ,\gamma_{l,k} } \right]^T},
	\end{equation}
	where 
	\begin{equation}\label{eta}
		{\eta_{l,k}} = {\left( {{\mathbf{w}^{2}_{l,k}} \cdot {\mathbf{w}^{1}_{l,k,1}}} \right)^T}g'_1({\mathbf{W}^{1}_{l,k}}\hat{\mathbf{x}}'_k + {\mathbf{b}^{1}_{l,k}}),
	\end{equation}
	\begin{equation}\label{gamma}
	\gamma_{l,k}=(\mathbf{w}^{2}_{l,k}\cdot\mathbf{w}^{1}_{l,k,2})^Tg'_1(\mathbf{W}^{1}_{l,k}\hat{\mathbf{x}}'_k+\mathbf{b}^{1}_{l,k}),
	\end{equation}
	which are Lines 11 - 12 of Algorithm 1. In the derivations, we use the property $g'_1(\cdot)=1-g_1(\cdot)^2$.}

{With indexes $l,l' \in \{1,2 \}$, the message $m_{f_{s_{l,k}}\rightarrow x_{l',k}}(x_{l',k}) $ is computed by the BP rule with the messages $m_{s_{l,k}\rightarrow f_{s_{l,k}}}(s_{l,k})$ and $\forall l''\ne l', {m_{x_{l'',k}\rightarrow f_{s_{l,k}}}(x_{l'',k})} $ later computed in (\ref{xlkslk}), yielding
	\begin{equation}\label{35}
		\begin{aligned}
			&m_{f_{s_{l,k}}\rightarrow x_{l',k}}(x_{l',k}) \\
			&= \langle f_{s_{l,k}}(s_{l,k},\mathbf{x}'_k) \rangle_{m_{s_{l,k}\rightarrow f_{s_{l,k}}}(s_{l,k}){m_{x_{l'',k}\rightarrow f_{s_{l,k}}}(x_{l'',k})}} \\
			&\propto \mathcal{N}( x_{l,k};{\psi}^{l,l'}_{l,k} ,{\tau}^{l,l'}_{{\psi}_{l,k}}),
		\end{aligned}
	\end{equation}
	where for $l'=1$
	\begin{equation}\label{5.9}
		{\tau}^{l,1}_{{\psi}_{l,k}}  = ({{\tau}_{q}}+\gamma_{l,k}^2{\tau}_{{x}_{2,k}})/{\eta}^2_{l,k}
	\end{equation}
	\begin{equation}\label{5.8}
		{\psi}^{l,1}_{l,k} =({q}_{l,k}-\tilde{q}_{l,k})/\eta_{l,k} +\hat{x}_{1,k}
	\end{equation}
	and for $l'=2$
	\begin{equation}\label{5.9_1}
		{\tau}^{l,2}_{{\psi}_{l,k}} = ({{\tau}_{q}}+\eta_{l,k}^2{\tau}_{{x}_{1,k}})/{\gamma}^2_{l,k}
	\end{equation}
	\begin{equation}\label{5.8_1}
		{\psi}^{l,2}_{l,k} =({q}_{l,k}-\tilde{q}_{l,k})/\gamma_{l,k} +\hat{x}_{2,k}
	\end{equation}
which are given in Lines 13 - 16 of Algorithm 1.
	The message $m_{x_{l,k}\rightarrow f_{x_{l,k}}}(x_{l,k})$ is calculated as
	\begin{equation}\label{Eq.29}
		\begin{aligned}
			n_{x_{l,k}\rightarrow f_{x_{l,k}}}(x_{l,k})&= {m_{f_{s_{1,k}}\rightarrow x_{l,k}}(x_{l,k})}{m_{f_{s_{2,k}}\rightarrow x_{l,k}}(x_{l,k})}  \\
			&\propto  \mathcal{N}(x_{l,k};{\psi}_{l,k},{\tau}_{{\psi}_{l,k}}),
		\end{aligned}
	\end{equation}
	with
	\begin{equation}\label{Eq.30}
		{\tau}_{{\psi}_{l,k}} = ( \frac{1}{{\tau}_{\psi_{l,k}}^{l,1}}+\frac{1}{{\tau}_{\psi_{l,k}}^{l,2}})^{-1},
	\end{equation}
	\begin{equation}\label{Eq.31}
		{\psi}_{l,k}= (\frac{{\psi}_{l,k}^{l,1}}{{\tau}_{\psi_{l,k}}^{l,1}}+\frac{{\psi}_{l,k}^{l,2}}{{\tau}_{\psi_{l,k}}^{l,2}}){\tau}_{{\psi}_{l,k}},
	\end{equation}
	which are given in Lines 17 and 18 of Algorithm 1.}

{Note that all the values in the above computations are real as the real parts and imaginary parts of the variables are separated.   
To facilitate the estimation of the complex-valued symbols, we merge the real and imaginary components. Hence, we have 
	\begin{equation}
       {\tau}_{\tilde{\psi}_{k}} ={\tau}_{{\psi}_{1,k}}+{\tau}_{{\psi}_{2,k}}
	\end{equation}
\begin{equation}
  \tilde{\psi}_{k} ={\psi}_{1,k}+j*{\psi}_{2,k},
\end{equation}
which are shown in Lines 19 and 20 of Algorithm 1.}

{The prior of $x_k$, which is a uniform discrete
distribution, i.e.,
\begin{equation}
	p(x_{k}=\lambda_a)=1/|\mathcal{A}|.
\end{equation}
It is not hard to show that the \textit{a posteriori} mean $\hat{x}_{k}$ and
variance $\tau_{x_{k}}$ of $x_{k}$ are given by (also shown in Lines 21 - 24 of Algorithm 1)
\begin{equation}\label{Postx}
\hat{x}_{k} =\sum^{|A|}_{a=1}\lambda_{a}\mu_{k,a} \end{equation}  
\begin{equation}
\tau_{x_{k}} =\sum^{|A|}_{a=1}\mu_{k,a}|\lambda_a-\hat{x}_k|^2 ,\end{equation}
where 
\begin{equation}
\mu_{k,a} = \xi_{k,a}/\sum^{|A|}_{a=1}\xi_{k,a} ,
\end{equation}  with
\begin{equation}
\xi_{k,a} = \text{exp}(-{\tau}_{\tilde{\psi}_{k} }^{-1}|\lambda_a-\tilde{\psi}_{k}|^2).
\end{equation}
To simplify the message computations, we use the following approximation: 
\begin{equation}\label{xlkslk}
	m_{x_{l,k}\rightarrow f_{s_{1,k}}}= m_{x_{l,k}\rightarrow f_{s_{2,k}}}=m_{f_{x_{k}}\rightarrow x_{l,k}}.
\end{equation}
Since the \textit{a posteriori} mean $\hat{x}_{k}$ of $x_{k}$ are updated in \eqref{Postx}, we update ${f}_l({\mathbf{x}'_k})$ (including $\tilde{q}_{l,k}$, $\eta_{l,k}$ and $\gamma_{l,k}$ ) in \eqref{flxk} with the updated  $\hat{x}_{k}$. This is Line 25 of Algorithm 1. 
}

{To compute the message $m_{f_{s_{l,k}}\rightarrow s_{l,k}}$, we separate the real part and imaginary part of ${x}_{k}$ and assume that they have the same variance, so
	\begin{equation}\label{tause}
		{\tau}_{{x}_{1,k}} ={\tau}_{{x}_{2,k}}=1/2\tau_{x_{k}}
	\end{equation}
	\begin{equation}\label{xse}
\hat{x}_{1,k}=\Re\{\hat{x}_{k}\},\hat{x}_{2,k}=\Im\{\hat{x}_{k}\},
\end{equation}
which are Lines 26 and 27 of Algorithm 1. 
	Then, we are ready to compute the message from $f_{s_{l,k}}$ to $s_{l,k}$, i.e.,
	\begin{equation}\label{lefts}
		\begin{aligned}
			m_{f_{s_{l,k}}\rightarrow s_{l,k}}(s_{l,k})&= \langle f_{s_{l,k}}(s_{l,k},\mathbf{x}'_k) \rangle_{\prod_{l'} m_{x_{l',k}\rightarrow f_{s_{l,k}}}(x_{l',k})} \\
			& \propto{\mathcal{N}(s_{l,k};\overleftarrow{s}_{l,k},\overleftarrow{\tau}_{s_{l,k}})},
		\end{aligned}
	\end{equation}
	with
	\begin{equation}\label{6.1}
		\overleftarrow{\tau}_{s_{l,k}}= \eta_{l,k}^2{\tau}_{{x}_{1,k}}+\gamma_{l,k}^2{\tau}_{{x}_{2,k}}
	\end{equation}
	\begin{equation}\label{6.2}
		\overleftarrow{s}_{l,k}=\tilde{q}_{l,k},
\end{equation}
which are Lines 28 and 29 of Algorithm 1. According to {UAMP version 2 \cite{9293406}}, an averaged variance is required, i.e.,
\begin{equation}
	\tau_s = \frac{1}{2K}\sum_{l=1}^{2}\sum_{k=1}^{2K}{\tau}_{s_{l,k}},
\end{equation}
which is Line 30 of Algorithm 1. This is the end of a single round iteration of the iterative process. A number of iterations can be performed until the algorithm converges, or the algorithm is terminated when a pre-set number of iterations is reached.
}
\section{Extension To Coded System With Turbo
Receiver}
\begin{figure}
	\begin{center}
		\includegraphics[width=3.5 in]{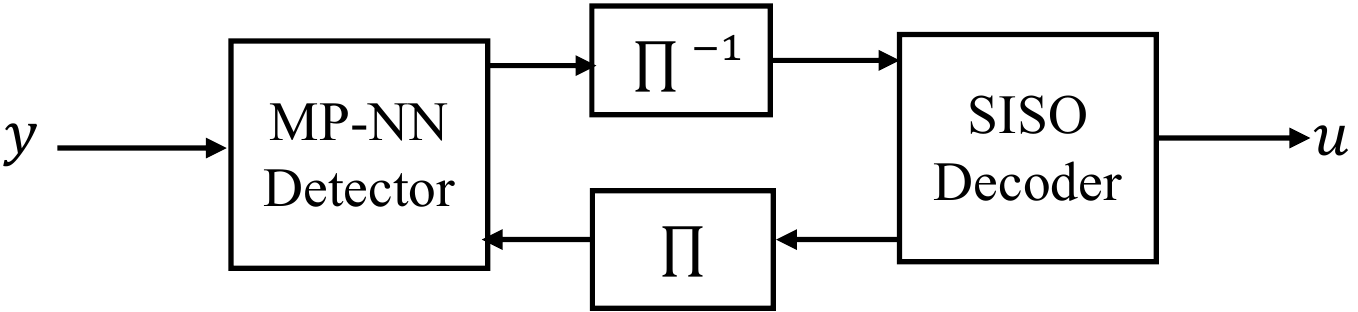}\\
		\caption{{ {Block diagram of turbo receiver, where $\Pi$ and $\Pi^{-1}$  denote an interleaver and the corresponding deinterleaver, respectively.}}}\label{coded}
	\end{center}
\end{figure}

{In a turbo receiver, the detector and decoder work in an iterative manner to achieve joint detection and decoding. It is well known that a turbo receiver can be much more powerful than a conventional non-iterative receiver {\cite{984761,4444762}}. Compared to the direct detectors, the proposed Bayesian detector can be readily extended to a SISO detector so that a turbo receiver can be implemented.}  In a turbo system, the information bits are firstly encoded and then interleaved before mapping. 
Each symbol $x_k \in \mathcal{A}=[\lambda_1,\ldots,\lambda_{|\mathcal{A}|}]$ 
is mapped from a sub-sequence of the coded bit sequence, which is denoted by $\mathbf{u}_k=[u_{k}^1,\ldots,u_{k}^{log|\mathcal{A}|}]$. Each $\lambda_a$ corresponds to a length-$log|\mathcal{A}|$ binary sequence denoted by $\{\lambda_a^1, \ldots, \lambda_a^{log|\mathcal{A}|} \}$.}

{{The turbo receiver is shown in Fig.~\ref{coded}, which consists of the UAMP-based Bayesian detector and a SISO decoder, working in an iterative manner to exchange extrinsic log-likelihood ratios (LLRs) of the coded bits. For simplicity, a single user is assumed in Fig.~\ref{coded}. The detector calculates the extrinsic LLRs for each coded bit with the extrinsic LLRs from the decoder as the \textit{a priori} information. Then, with the extrinsic LLRs from the detector, the decoder refines the LLRs with the code constraints. In this work, we assume a standard SISO decoder (e.g., the Bahl–Cocke–Jelinek–Raviv (BCJR) algorithm for convolutional codes) is employed, and we adapt the detector proposed in Section III to a SISO one.}

{The task of the detector is to calculate the extrinsic LLR for each code bit $u_{k}^q(m)$, which can be represented as
\begin{align}
    	{L}^e(u_{k}^q) &  = \ln{\frac{p(u_{k}^q=0\lvert\mathbf{r})}{p(u_{k}^q=1\lvert\mathbf{r})}}-L^a(u_{k}^q),       
\end{align}
where $L^a(u_{k}^q)$ is the output extrinsic LLR of the decoder in the previous iteration. The extrinsic LLR ${L}^e(u_{k}^q)$ is passed to the decoder. The derivation for ${L}^e(u_{k}^q)$ in terms of extrinsic mean and variance can be found in \cite{5741768}, and ${L}^e(u_{k}^q)$  can be expressed as 
\begin{equation}\label{LLRE}
    {L}^e(u_{k}^q) = \ln \frac{\sum\limits_{\lambda_a \in \mathcal{A}^0_{q}}\exp(-\frac{|\lambda_a-m_{x_{k}}^e|^2}{v^e_{k}})\prod\limits_{q'\neq q}p(u^{q'}_{k}= \lambda_a^{q'})}{\sum\limits_{\lambda_a \in \mathcal{A}^1_{q}}\exp(-\frac{|\lambda_a-m_{x_{k}}^e|^2}{v^e_{k}})\prod\limits_{q'\neq q}p(u^{q'}_{k}= \lambda_a^{q'})},
\end{equation}
where $\mathcal{A}_q^0$ and $\mathcal{A}_q^1$ denote subsets of all $\lambda_a\in D$ whose label in position $q$ has the value of 0 and 1, respectively, and $m_{k}^e$ and $v^e_{k}$ are the extrinsic mean and variance of $x_{k}$. According to \cite{5741768}, the extrinsic variance and mean are defined as  
\begin{equation}
v_{k}^e =(\frac{1}{{v}^p_{k}}-\frac{1}{v_{k}})^{-1}
\end{equation}
\begin{equation}
m_{x_{k}}^e =v_{x_{k}}^e (\frac{m^p_{x_{k}}}{{v}^p_{{x_{k}}}}-\frac{m_{x_{k}}}{v_{{x_{k}}}}),
\end{equation}
where $m_{x_{k}}$ and $v_{{x_{k}}}$ are the \textit{a priori} mean and variance of $x_{k}$ calculated based on the output LLRs of the SISO decoder \cite{984761,4444762,vucetic2012turbo}, and $m^p_{x_{k}}$ and ${v}^p_{{x_{k}}}$ are the \textit{a posteriori} mean and variance of $x_{k}$.} {By examining the derivation of the Bayesian detector in Algorithm 1, we can find that $\tilde{\psi}_{k}$ and ${\tau}_{\tilde{\psi}_{k}}$ consist of the extrinsic mean and variance of ${x_{k}}$ as they are the messages passed from observation and do not contain the immediate \textit{a priori} information about $x_{k}$. Therefore, we have
\begin{eqnarray}
m_{x_{k}}^e = \tilde{\psi}_{k}, ~~
v_{x_{k}}^e = \tau_{\tilde{\psi}_{k}}.
\end{eqnarray}
Then, (\ref{LLRE}) can be readily used to calculate the extrinsic LLRs of the coded bits. 
Note that with the LLRs output from the SISO decoder, we can compute the probability $p(x_{k} = \lambda_a)$ for each $x_{k}$, which is no longer $1/|\mathcal{A}|$ in Algorithm 1. Therefore, $\xi_{k,a}$ in Line 21  of Algorithm 1 needs to be changed to 
\begin{equation}
    \xi_{k,a} = p(x_{k}=\lambda_a)\exp(-v^{-1}_{\psi_k}|\lambda_a-\tilde{\psi}_{k}|^2).
\end{equation}
In addition, we note that the iteration of the detector can be incorporated into the iteration between the SISO decoder and detector, i.e., only a single loop iteration (without inner iteration) is needed.
}

\section{Simulation Results}

\begin{figure}
	\begin{center}
		\includegraphics[width=3.5in]{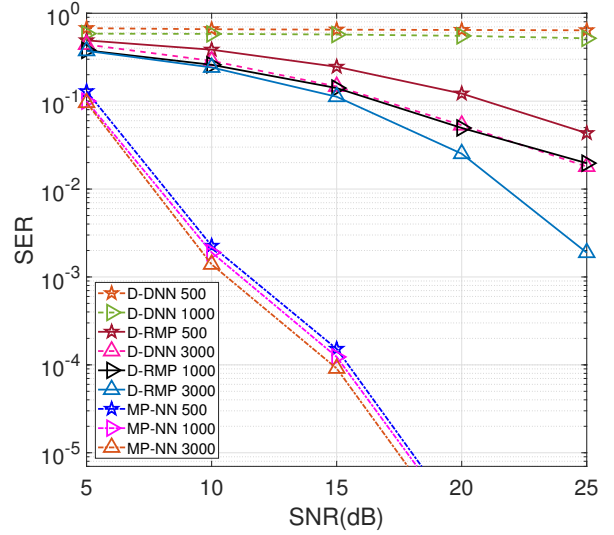}\\
		\caption{{SER performance comparisons of MP-NN, D-DNN and D-RMP based detectors with different training lengths.}}\label{SERTaininglength}
	\end{center}
\end{figure}

Assume that the BS is equipped with a uniform linear antenna array, $N=10$ and $K=5$. The modulation scheme used is 16-QAM. {The Saleh-Valenzuela channel model \cite{6847111} is employed. The channel vector $\mathbf{h}_k$ between the $k$th user and the $N$ receive antennas is represented as
\begin{equation}\label{vmodel}
    \mathbf{h}_k=\sqrt{\frac{N}{Q_k}}\sum^{Q_k}_{q=1}\beta_{kq}\mathbf{a}(\theta_{kq}),
\end{equation}
where $\theta_{kq}$ is the incident angle of the $q$th path, $\mathbf{a}(\theta_{kq}) = \frac{1}{\sqrt{N}}[1, e^{-j2\pi d sin(\theta_{kq})/ \lambda},\ldots,e^{-j2\pi d sin(\theta_{kq})(N-1)/ \lambda}]^T$ is a length-$N$ steering vector with antenna spacing $d$, $\lambda$ is the wavelength of carrier,  $Q_k$ is the number of paths for user $k$, and $\beta_{kq}$ is the complex gain of the $q$th path.} We use the same parameter settings as in \cite{6831645}, where $d=\lambda/2$, $Q = 3$, $\beta_{kq}$ follows Gaussian distribution with zero mean and unity variance, 
and $\theta_{kq}$ is uniformly drawn from $(-0.5 \pi, 0.5\pi]$.
{As in \cite{7058359,ieee802}, the parameters used for the PA nonlinearity are $\alpha_a = 4.65$, $\alpha_{\phi} = 2560$, $\beta_{\phi} = 0.114$, $\sigma_a = 0.81$, $x_{\text{sat}} = 0.58$, $q_1 = 2.4$ and $q_2 = 2.3$.}  {For I/Q imbalance, the parameters are $\theta_k=4^{\circ}$ } and $\lambda_k=0.05$.
The SNR is defined as $P_x/\sigma^2_n$, where $P_x$ is the power of the transmitted signal of a user (assuming all users have the same transmit power), and $\sigma^2_n$ is the power of the noise (per receive antenna) at the receiver. {We compare the proposed detector called MP-NN, where the parameters of
the sub-NN pairs are tied, with existing detectors, including DNN based direct detector \cite{8476222} and RMP-based direct detector \cite{7600411}, which are called D-DNN and D-RMP, respectively.}

\begin{figure}

	\begin{center}
		\includegraphics[width=3.5in]{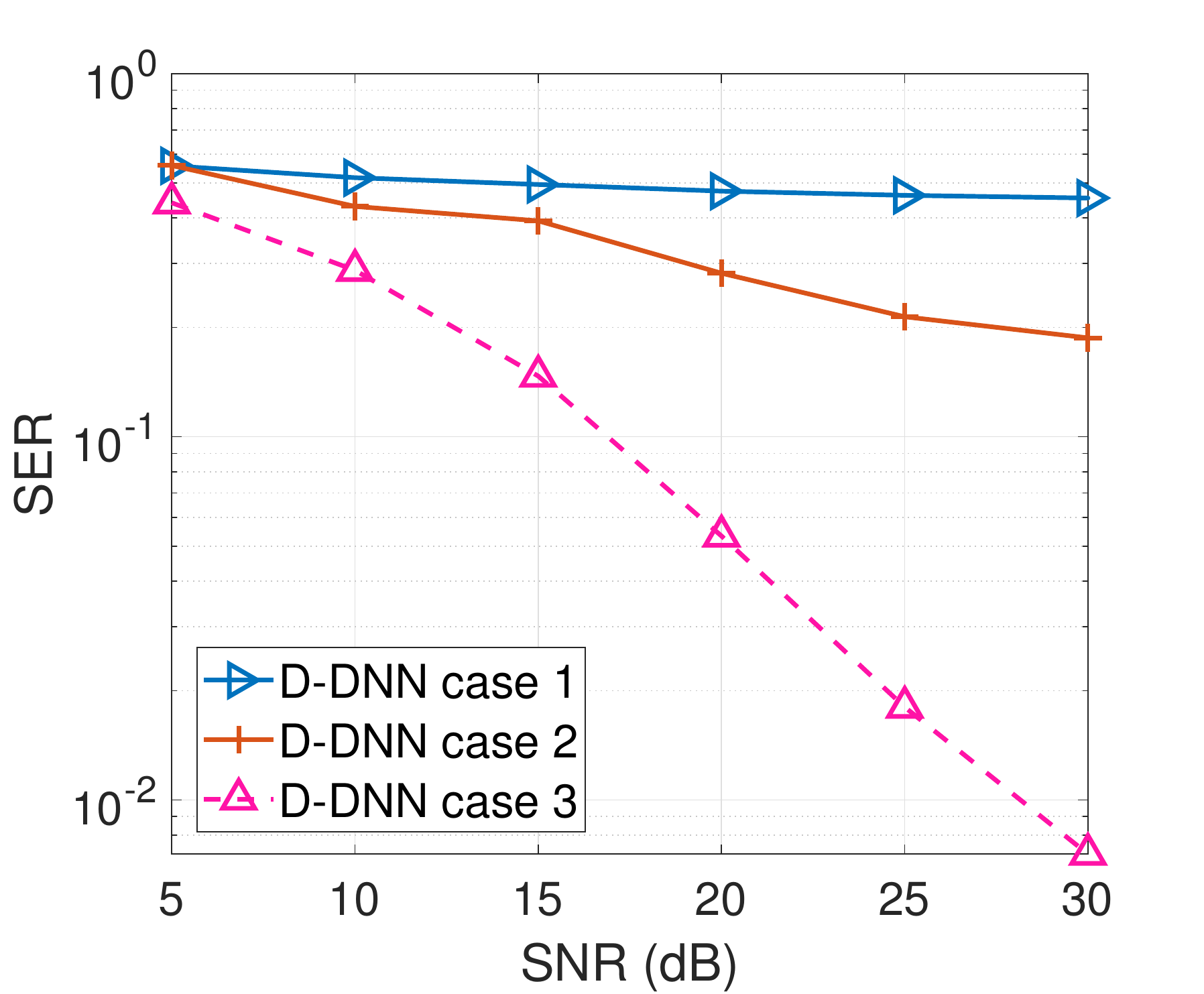}\\
		\caption{{SER performance comparisons of D-DNN with different hyper-parameters.}}\label{DNNexample}
	\end{center}
\end{figure}

The deep learning framework \textit{Tensorflow} is used for {(D)NN training and validation}. Batch gradient descent is adopted, and cross-validation is used to avoid overfitting and ensure the generality of the trained model. We use 80\% and 20\% of the dataset for training (including 3-fold validation) and testing.} Through the validation data set, we determine that the batch size is 100 and the number of epochs is 300. The Adam optimizer with a learning rate 0.01 is employed to update the (D)NN parameters. For the proposed NN, the number of hidden nodes $N'$ in the sub-NNs is $20$. For D-DNN, the activation function $Tanh$ is employed for hidden layers. The number of hidden layers is 2, and the numbers of nodes of the hidden layers are 30 and 40, unless these parameters are specified. For D-RMP, a fifth order polynomial is employed. 

\subsection{Uncoded System} 

{We first consider a uncoded system. Fig. \ref{SERTaininglength} shows the symbol error rate (SER) of the detectors, where the training lengths 500, 1000 and 3000 are used to examine the impact of training length on the performance of the detectors. From the results, we can see that in all the cases, the proposed MP-NN detector always performs remarkably better than other detectors. We can also see that D-RMP performs better than D-DNN. Moreover, when the training length is decreased from 3000 to 500, there are only minor changes in the performance of MP-NN, which indicates that the training length 500 is sufficient for MP-NN. In contrast, the impact of the training length on the performance of D-RMP and D-DNN is significant, and their performance degrades rapidly with the reduce of the training length. These results demonstrate the effectiveness of the proposed detector, i.e., it can be trained more effectively and the Bayesian detector is much more powerful. Considering that neither D-RMP nor D-DNN works well with training lengths 500 and 1000, we use training length 3000 in the subsequent simulations. }

{With the training length fixed to 3000, we examine the performance of D-DNN by changing its hyper-parameters including the number of layers and hidden nodes, which are indicated by cases 1, 2 and 3. In case 1, the number of hidden layers is 2, we increase the number of hidden nodes in the two hidden layers to 300 and 400, respectively. In case 2, we increase the number of hidden layers to 3 with hidden nodes 30, 40 and 50, respectively. In case 3, we use the default setup as before. The results are shown in Fig. \ref{DNNexample}. It can be seen that, compared to the default hyper-parameter setting (case 3), the SER performance of D-DNN deteriorates significantly with other settings. This is because the number of parameters for the DNN is increased significantly in cases 1 and 2, and the training samples are insufficient. Hence in the subsequent examples, we will use the the default setting for D-DNN.}   

\begin{figure}
	\begin{center}
		\includegraphics[width=3.5 in]{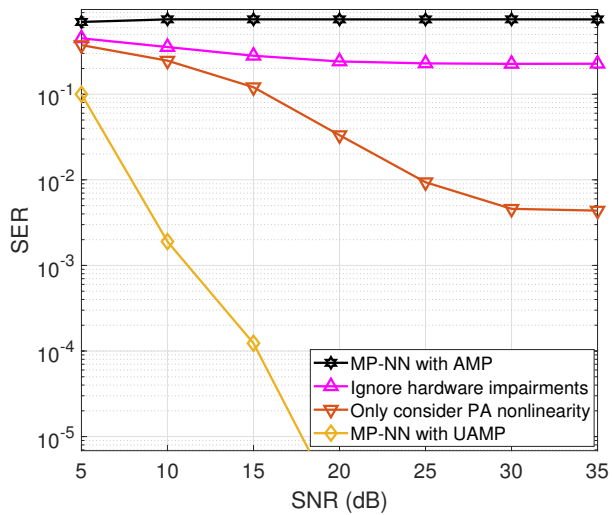}\\
		\caption{{SER performance of the MP-NN detector, the receiver without handling I/Q imbalance, and the receiver without handling  nonlinearity and I/Q imbalance. }}\label{UAMPAMP}
	\end{center}
\end{figure}

{It is mentioned in the previous section that, UAMP plays a crucial role in the {message passing based Bayesian detector MP-NN}. To demonstrate this, we also use AMP to deal with the densely connected part of the factor graph (i.e., AMP is integrated into the message passing algorithm). We compare the SER performance of the detector with AMP and UAMP in Fig. \ref{UAMPAMP}. We can see that the AMP based detector simply does not work as the AMP algorithm does not converge. 
To demonstrate that it is necessary to handle the I/Q imbalance and PA nonlinearity at the receiver side, we compare the MP-NN receiver with  
the receiver without considering I/Q imbalance and nonlinearity, where the zero-forcing (ZF) detector with known MIMO channel matrix is employed. We also compared the proposed receiver with the receiver without considering I/Q imbalance, where polynomial based detector is employed to handle PA nonlinearity. 
The results are also shown in Fig. \ref{UAMPAMP}. It can be seen that, without considering both I/Q imbalance and PA nonlinearity, the receiver simply does not work properly. If only PA nonlinearity is considered, the receiver performs poorly and a very high SER floor is observed. The results indicate that both I/Q imbalance and PA nonlinearity need to be properly handled by the receiver to achieve good performance.  } 

\begin{figure}
	\begin{center}
		\includegraphics[width=3.5in]{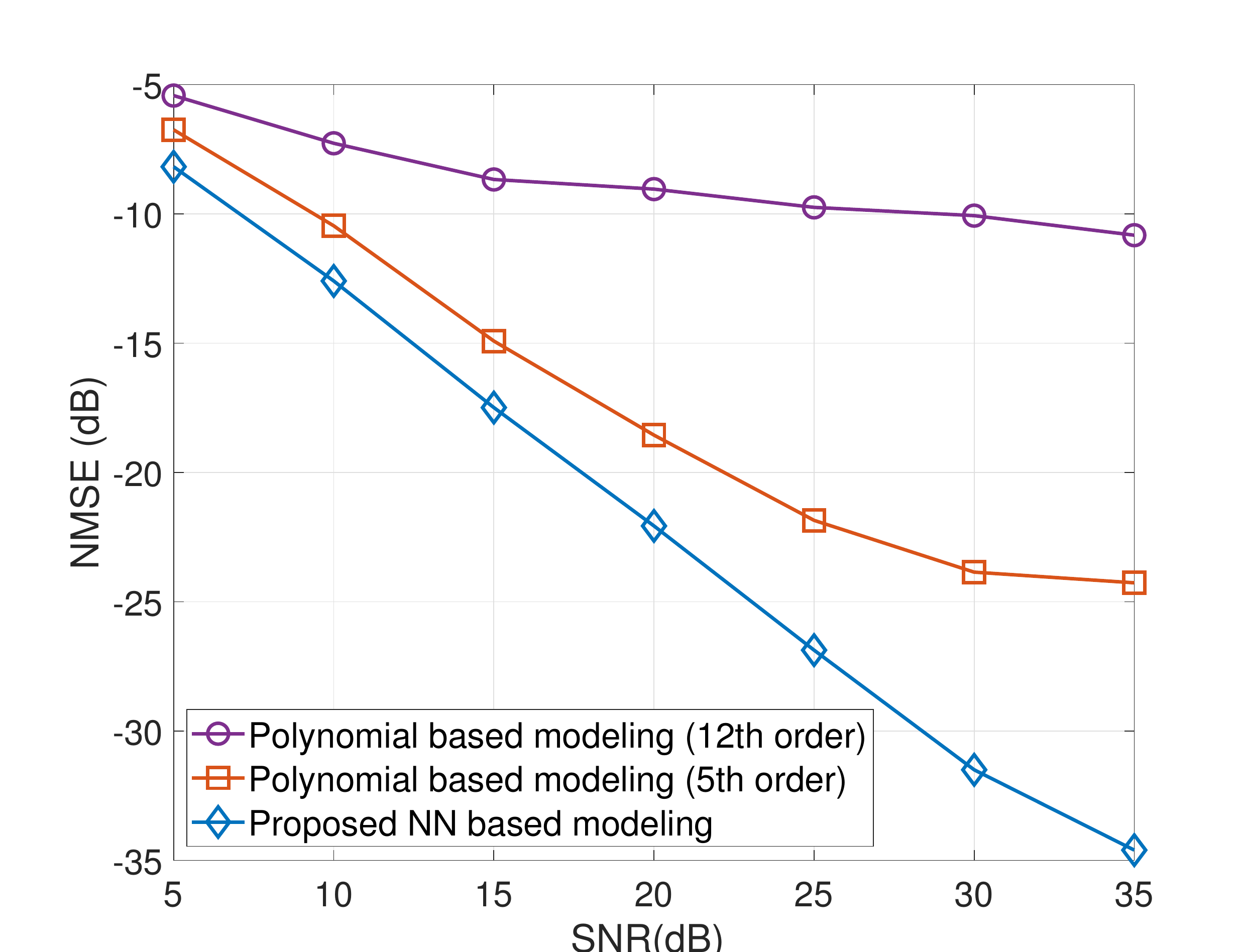}\\
		\caption{{Modeling performance comparison of the proposed NN and polynomial methods with $5$th and $12$th orders, respectively. }}\label{Modeling}
	\end{center}
\end{figure}

\begin{figure}
	\begin{center}
		\includegraphics[width=3.5in]{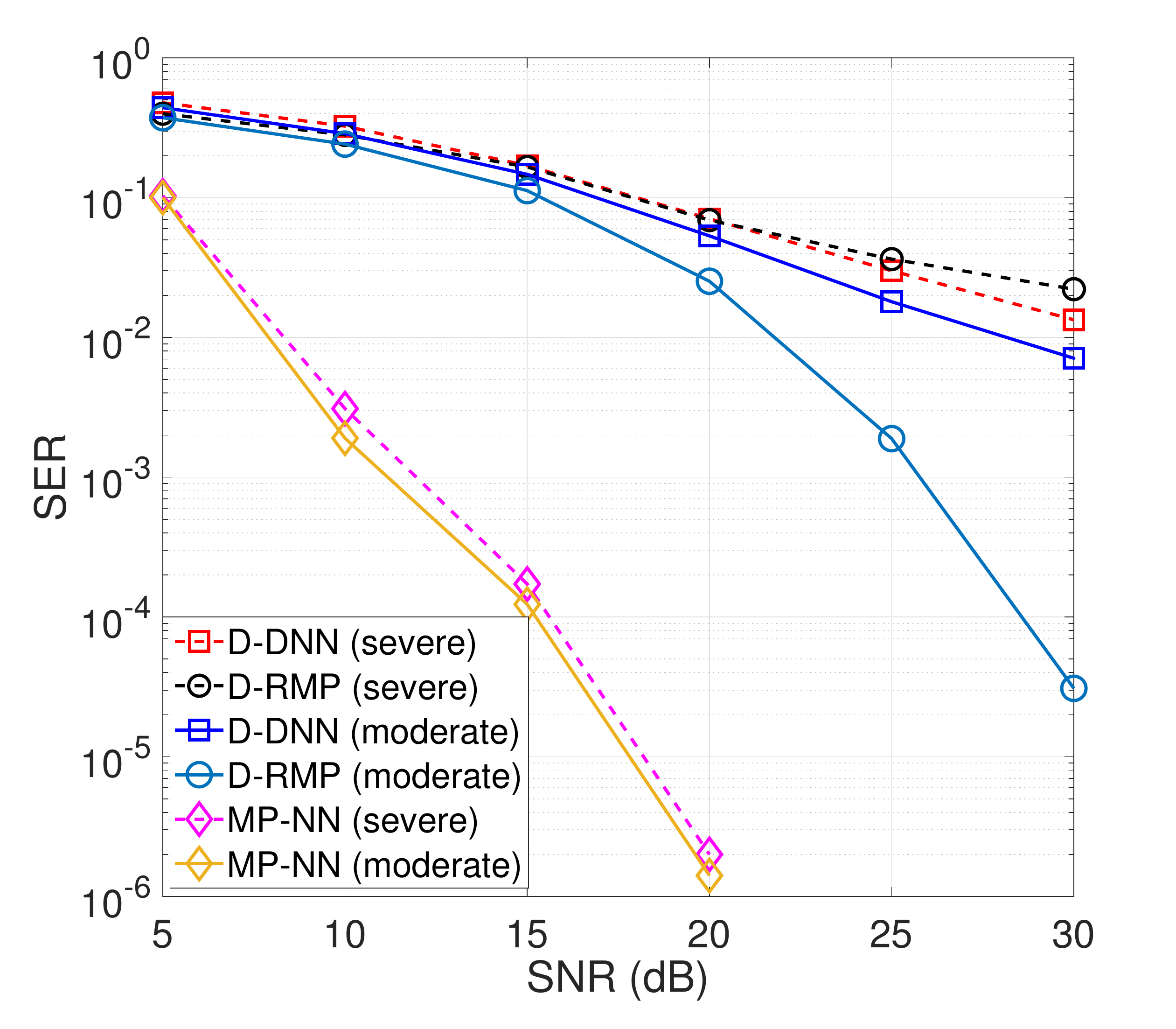}\\
		\caption{{SER performance comparisons of the receivers with moderate and severe hardware imperfections.}}\label{hardwarechange}
	\end{center}
\end{figure}

As discussed in the previous section, the architecture of the NN proposed in this work is designed based on signal flow to model the joint effects of hardware impairments and co-channel interference. We note that the polynomial techniques \cite{7600411} can also be used to model the joint effects. It is interesting to compare the performance of the two methods. We use the normalized mean square error (NMSE) to evaluate the modelling performance and the results are shown in Fig. \ref{Modeling}, {where polynomials with the 5th and 12th order are used. We note that, although the use of higher order polynomial may improve the modelling capability of the polynomial technique, it causes difficulties in determining the polynomial parameters due to numerical instability. {Moreover, it is noted that when the order of polynomial increases by one, the number of parameters to be determined is increased by $4KN(L+1)$, which is a significant increase, making it prone to overfitting due to the limited number of training samples.} 
As shown in Fig. \ref{Modeling}, the proposed NN significantly outperforms the 5th-order polynomial, indicating that the proposed NN has much better modeling capability. When increasing the polynomial order to 12, the performance of the polynomial method becomes extremely poor due to numerical instability and overfitting. The results demonstrate the advantage of the proposed NN in modelling.}


\begin{figure}
	\begin{center}
		\includegraphics[width=3.5in]{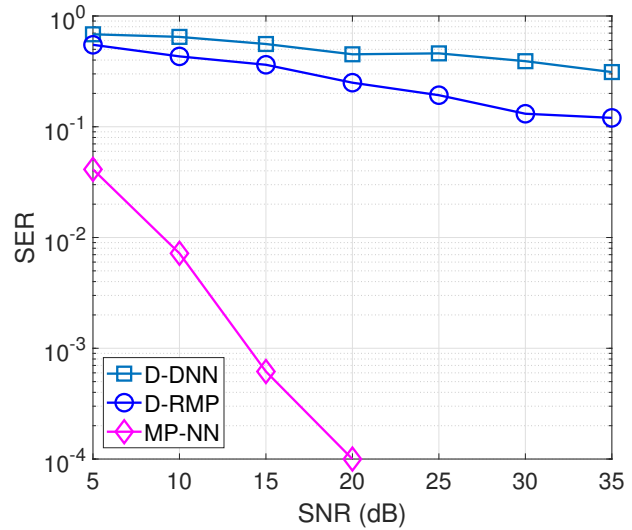}\\
		\caption{{SER performance comparison with extreme I/Q imbalance and PA nonlinear distortion.}}\label{extremecondi}
	\end{center}
\end{figure}


 So far, we have compared the performance of the receivers with moderate hardware imperfections. It is also interesting to test the capabilities of the receivers in handling severer hardware imperfections. According to \cite{9324825}, we increase the gain $\alpha_a$ of amplitude to amplitude conversion to 6.5 to simulate severer PA nonlinearity.  Fig. \ref{hardwarechange} shows the SER performance of the receivers. 
It can be seen that the performance of D-RMP and D-DDN deteriorate significantly with severer hardware imperfections. In contrast, the proposed MP-NN receiver only incurs marginal performance loss, and it still delivers outstanding performance. {We also adjust the I/Q imbalance and PA nonlinearity to an extreme condition. The PA nonlinearity is simulated using a fifth-order polynomial in \cite{7492233}. The I/Q imbalance parameter $\theta_k$ is increased to  $10^{\circ}$. The results are shown in Fig.~\ref{extremecondi}, where we can see that D-RMP and D-DNN simply do not work under the extreme hardware imperfections. In contrast, the proposed MP-NN detector still performs very well. These results demonstrate the high capability of MP-NN to deal with hardware distortions.}

\begin{figure}
	\begin{center}
		\includegraphics[width=3.5in]{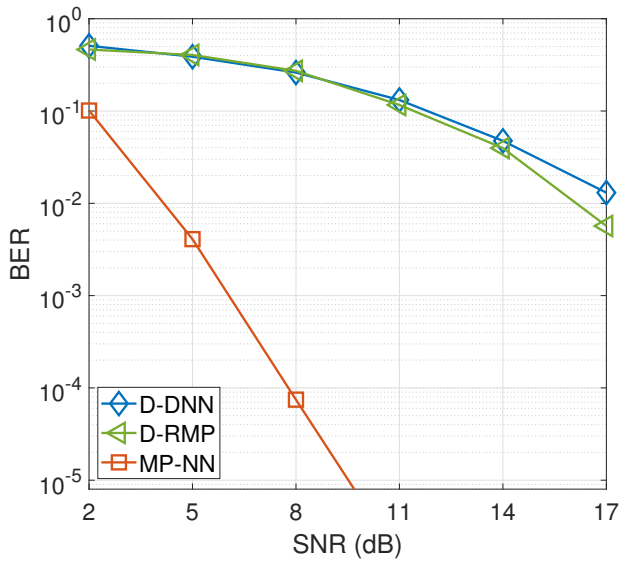}\\
		\caption{{BER performance of MP-NN, D-RMP and D-DNN receivers in a coded system.}}\label{Fig11}
	\end{center}
\end{figure}

\subsection{Coded System}
{ We then evaluate the performance of the detectors in a coded system, and compare the performance of the systems with and without turbo receiver. We use a rate-2/3 convolutional code with
generators [23, 35], followed by a random interleaver and 16-QAM modulation, where Gray mapping is used in symbol mapping. The BCJR algorithm is used to implement the SISO decoder. As it is unknown how to implement a turbo receiver based on the direct detectors D-RMP and D-DNN, so non-iterative receivers are implemented for them, where the outputs of detectors after hard decision are fed to a Viterbi decoder. The other settings are the same as those in the previous section, and the bit error rate (BER) is used to evaluate the performance of the receivers.}
{
{We  compare the performance of the MP-NN turbo receiver, D-RMP receiver and D-DNN receiver in the coded system. Fig.~\ref{Fig11} shows the BER performance of the receivers.   
We can see that the proposed MP-NN detector performs significantly better than other receivers. Similar to the previous results, the D-RMP receiver performs slightly better than the D-DNN receiver.}  

\section{Conclusions}
{In this work, we developed a Bayesian detector for MIMO communications with combined hardware imperfections. Based on the signal flow, we first design the architecture of an NN to model the hardware imperfections and multi-user interference, so that the NN can be trained much more efficiently, compared to conventional DNN-based methods. Then, representing the trained NN as a factor graph and leveraging UAMP, we develop an efficient message passing based Bayesian detector MP-NN. Both non-iterative receiver and turbo receiver are investigated. Extensive simulation results demonstrate that the proposed method significantly outperforms state-of-the-art methods.}

\rev{By combining NN and factor graph techniques, this work provides a general way to achieve Bayesian signal detection for a communication system with complicated input-output relationship. Interestingly, a recent work in \cite{newadd} also combines NNs and factor graphs for stationary time sequence inference. However, the ways of combining NNs and factor graphs in this work and \cite{newadd} are very different. Here, NNs are represented as factor graphs to develop efficient message passing algorithms for Bayesian inference, where message passing is carried out on NNs. In \cite{newadd}, NNs are used to learn specific components of
a factor graph describing the distribution of the time sequence, where NNs are involved in the computation of local messages. Combining NNs and factor graphs is promising to tackle challenging signal processing tasks, which is worth further exploration.}

\ifCLASSOPTIONcaptionsoff
  \newpage
\fi



\normalem
\bibliographystyle{IEEEtran}
\bibliography{IEEEabrv,bare_jrnl_comsoc}
\end{document}